\documentclass[12pt, draftclsnofoot, onecolumn]{IEEEtran}
 \usepackage{amsmath,amssymb}
 \usepackage{subfigure}
 \usepackage{graphicx,graphics,color,psfrag}
 \usepackage{cite,balance}
 \usepackage{caption}
 \allowdisplaybreaks
 \usepackage{algorithm}
 \usepackage{accents}
 \usepackage{amsthm}
 \usepackage{bm}
 \usepackage{algorithmic}
 \usepackage[english]{babel}
 \usepackage{multirow}
 \usepackage{enumerate}
 \usepackage{cases}
 \usepackage{stfloats}
 \usepackage{dsfont}
 \usepackage{color,soul}
 \usepackage{amsfonts}
 \usepackage{cite,graphicx,amsmath,amssymb}
 \usepackage{subfigure}
 \usepackage{fancyhdr}
 \usepackage{hhline}
 \usepackage{graphicx,graphics}
 \usepackage{array,color}

 \usepackage{amsmath}
 
 \usepackage{lipsum}
\usepackage{mwe}

\makeatletter
\newenvironment{subtheorem}[1]{%
  \def\subtheoremcounter{#1}%
  \refstepcounter{#1}%
  \protected@edef\theparentnumber{\csname the#1\endcsname}%
  \setcounter{parentnumber}{\value{#1}}%
  \setcounter{#1}{0}%
  \expandafter\def\csname the#1\endcsname{\theparentnumber.\Alph{#1}}%
  \ignorespaces
}{%
  \setcounter{\subtheoremcounter}{\value{parentnumber}}%
  \ignorespacesafterend
}
\makeatother
\newcounter{parentnumber}

\newtheorem{theorem}{Theorem}

\newtheorem{definition}{Definition}

\newtheorem{lemma}{Lemma}

\newtheorem{proposition}{Proposition}

\newtheorem{assumption}{Assumption}

\newtheorem{remark}{\bf Remark}

\def\E{\mathsf{E}}

\def\phi{\varphi}

\def\SIR{\mathsf{SIR}}

\def\l{\left}
\def\r{\right}
\def\({\left(}
\def\){\right)}

\def\b0{{\mathbf{0}}}

\newcommand{\nn}{\nonumber}

\begin{document}

\title{\huge Wireless Networks for Mobile Edge Computing: \\ \vspace{-5pt} Spatial Modeling and Latency Analysis}

\author{\vspace{-10pt}Seung-Woo Ko, Kaifeng Han,  and Kaibin Huang\\
\setlength{\skip\footins}{0.1cm}\thanks{ S.-W. Ko, K. Han and K. Huang are with 
The  University of  Hong Kong, Hong Kong (Email: 
huangkb@eee.hku.hk).}
}
\maketitle

\vspace{-60pt}
\begin{abstract}
\vspace{-10pt}


Next-generation wireless networks will provide users ubiquitous low-latency computing services using devices  at the network edge, called \emph{mobile edge computing} (MEC). The  key operation  of MEC is to offload computation intensive tasks from users.  Since each edge device comprises an \emph{access point} (AP) and a  \emph{computer server} (CS), a MEC network can be decomposed as a radio-access network cascaded with a CS network.  Based on the   architecture, we investigate network-constrained latency performance, namely communication latency  and computation latency under the constraints of radio-access  coverage and CS stability. To this end, a spatial random network  is modelled featuring random node distribution, parallel computing, non-orthogonal multiple access, and random computation-task generation. Given the model and  the said network  constraints, we derive the scaling laws of communication latency and computation latency with respect to network-load parameters (density of mobiles and their task-generation rates) and network-resource parameters (bandwidth, density of APs/CSs, CS computation rate).  
Essentially,  the analysis  involves the interplay of theories of  stochastic geometry, queueing, and parallel computing.  Combining the  derived scaling laws quantifies the tradeoffs between the latencies, network coverage and network stability. The results provide useful guidelines for MEC-network provisioning and planning  by avoiding either of the cascaded radio-access  network or  CS network  being a performance bottleneck.

\end{abstract}
\vspace{-20pt}
\section{Introduction}\label{Introduction}

One key mission of 5G systems is to provide users ubiquitous computing services (e.g., multimedia processing,  gaming and augmented reality)  using servers at the network edge, called \emph{mobile edge computing} (MEC) \cite{patel2014mobile}. Compared with cloud computing,  MEC can dramatically reduce  latency by avoiding transmissions over the backhaul network,  among many other advantages such as security and context awareness \cite{taleb2017multi, Mao:2017mec}. Most existing work focuses on designing MEC techniques by merging two disciplines: wireless communications and mobile  computing. In this work, we explore a different direction, namely the design of large-scale MEC networks with infinite nodes. To this end, a model of MEC network  is constructed featuring  spatial  random distribution of network nodes, wireless transmissions, parallel computing at servers. Based on the model and under network performance constraints, the latencies for communication and computation are analyzed by applying theories of stochastic geometry, queueing, and parallel computing. The results yield useful guidelines for MEC network provisioning and planning.

\vspace{-15pt}
\subsection{Mobile Edge Computing}
\vspace{-5pt}
To realize the vision of \emph{Internet-of-Things} (IoT) and smart cities, MEC is a key enabler providing ubiquitous and low latency access to computing resources. 
Edge servers in proximity of users are able to process  a large volume of data collected from IoT sensors and provide intelligent real-time solutions for various applications, e.g., health care,  smart grid, and autonomous driving. 
Due to its promising potential and the interdisciplinary nature, many new research issues arise in the area of MEC and are widely studied in different fields (see e.g., \cite{Mao:2017mec, Abbas2017, Mach2017}).

In the area of MEC, one research thrust focuses on designing techniques for enabling low-latency and energy-efficient \emph{mobile computation offloading} (MCO), which offloads computation intensive tasks from mobiles to the edge servers  \cite{Zhang2013TVT, Kwak2015, mao2016dynamic, you2016energy1,  kao2017hermes, liu2016delay, m2016optimal, ko2017live}. In \cite{Zhang2013TVT}, considering a CPU with a controllable clock, the optimal policy is derived using stochastic-optimization theory for jointly controlling the MCO decision (offload or not) and clock frequency with the objective of minimum mobile energy consumption. A similar design problem is tackled in \cite{Kwak2015} using a different approach based on Lyapunov optimization theory. Besides  MCO, the battery lives of mobile devices can be further lengthened by  energy harvesting  \cite{mao2016dynamic} or wireless power transfer  \cite{you2016energy1}.   The optimal policies for MEC control are more complex as they need to account for energy randomness  \cite{mao2016dynamic} or adapt the operation modes (power transfer or offloading) \cite{you2016energy1}. Designing energy-efficient MEC techniques under computation-deadline constraints implicitly attempts to optimize the latency-and-energy tradeoff. The problem of optimizing this tradeoff via computation-task scheduling is formulated  explicitly in \cite{kao2017hermes} and \cite{liu2016delay}  and solved using optimization theory. In addition, other  design issues for MEC are also investigated in the literature such as optimal program partitioning for partial offloading    \cite{m2016optimal} and data prefetching based on computation prediction  \cite{ko2017live}. 

Recent research in MEC focuses on designing more complex    MEC systems for multiuser MCO \cite{you2017energyMEC, sardellitti2015joint, chen2016efficient, kaewpuang2013framework,  rimal2016mobile, taleb2016follow, farris2017providing}. One important issue is the joint radio-and-computation resource allocation for minimizing sum mobile energy consumption under their deadline constraints.
The problem  is challenging due to the multiplicity of parameters and constraints involved in the problem including  multi-user channel states, computation capacities of servers and mobiles, and individual deadline and power constraints.  A tractable approach for solving the problem is developed in  \cite{you2017energyMEC} for a single-cell system comprising one edge server for multiple users. Specifically,  a so-called \emph{offloading priority function} is derived that includes all the parameters and used to show a simple threshold based structure of  the optimal policy. The problem of joint resource allocation in multi-cell systems is further complicated by the existence of inter-cell interference. An attempt is made in \cite{sardellitti2015joint} to tackle this problem using optimization theory. In distributed systems without coordination, mobiles make individual offloading decisions. For such systems, it is proposed in \cite{chen2016efficient} that  game theory is applied to improve the performance of distributed joint resource allocation in terms of latency and mobile energy consumption.

Cooperation between edge servers (or edge clouds) allows their resource pooling and sharing, which helps overcome their limitations in computation capacity. Algorithms for  edge-cloud cooperation are designed in  \cite{kaewpuang2013framework} based on game theory that enables or disables cooperation so as to  maximize the revenues of edge clouds  under the constraint of meeting  mobiles' computation demands. Compared with the edge cloud, the central cloud has unlimited computation capacity but its long distance from users can incur long latency for  offloading. Nevertheless, cooperation between edge and central clouds is desirable when the formers are overloaded. Given such cooperation, queueing theory is applied in \cite{rimal2016mobile} to analyze the latency for computation offloading. On the other hand,  cooperation between edge clouds   can support mobility by migrating computation tasks between servers. Building on the migration technology, a  MEC framework for supporting mobility is proposed in \cite{taleb2016follow} to adapt the placements of offloaded tasks in the cloud infrastructure depending on the mobility of the task owners. Besides offloaded tasks, computing services can be also migrated to adapt to mobility but service migration can place a heavy burden on the backhaul network or result in excessive latency. To address this issue, the framework of service duplication by virtualization is proposed in \cite{farris2017providing}.

\begin{figure}[t]
\centering
\centering
\includegraphics[width=7cm]{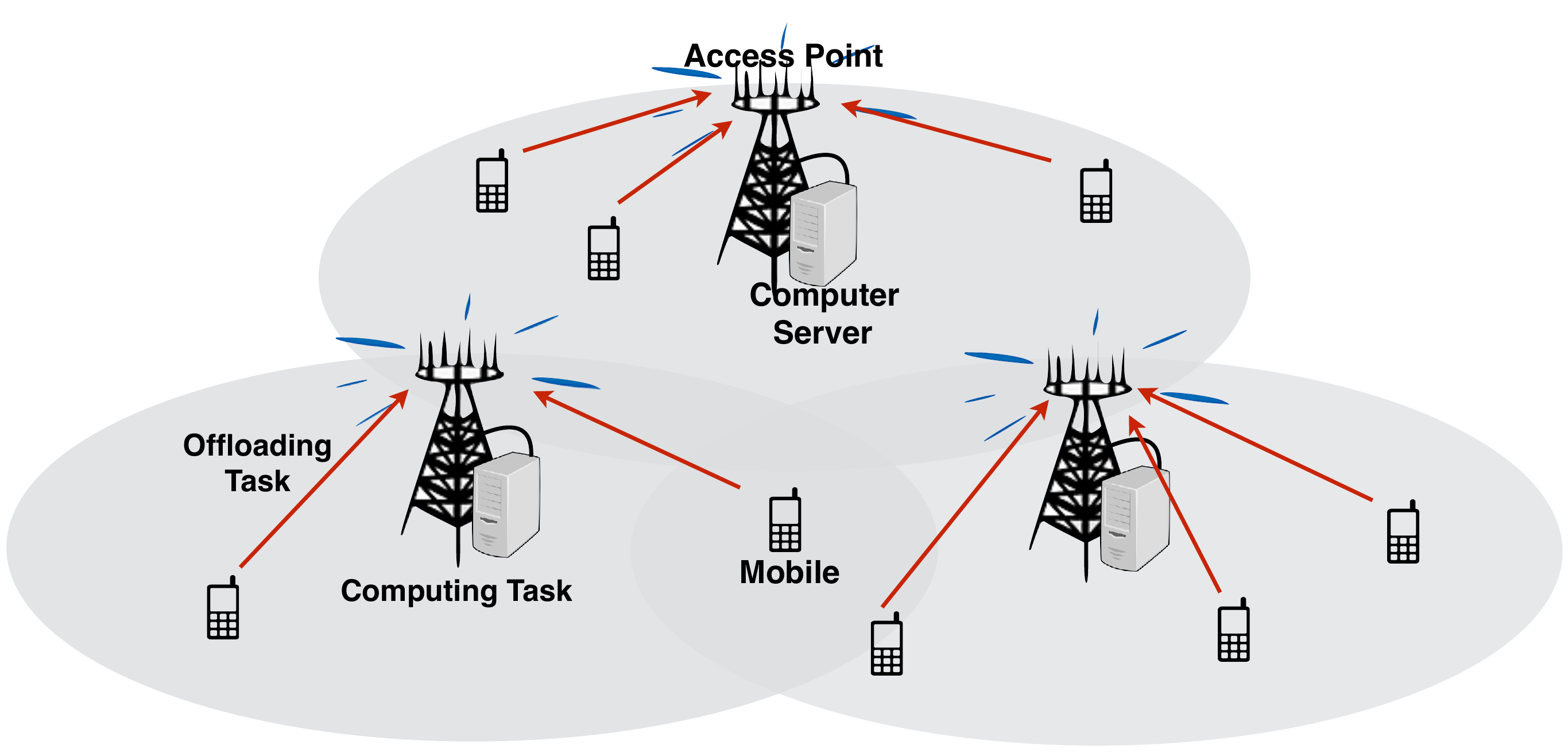}
\vspace{-10pt}
\caption{A MEC network where mobiles offload computation tasks to \emph{computer servers} (CSs) by wireless transmission to \emph{access points} (APs). }\label{MEC_system}
\vspace{-20pt}
\end{figure}

Prior work considers small-scale MEC systems with several users and servers/clouds, allowing the research to focus on designing complex MCO techniques and protocols. On the other hand, it is also important to study a large-scale MEC network with infinite nodes as illustrated in Fig.~\ref{MEC_system}, which is an area not yet explored. From the practical perspective, such studies can yield guidelines and insights useful for operators'  provisioning and planning of MEC networks.

\begin{figure}[t]
\centering
{\includegraphics[width=10cm]{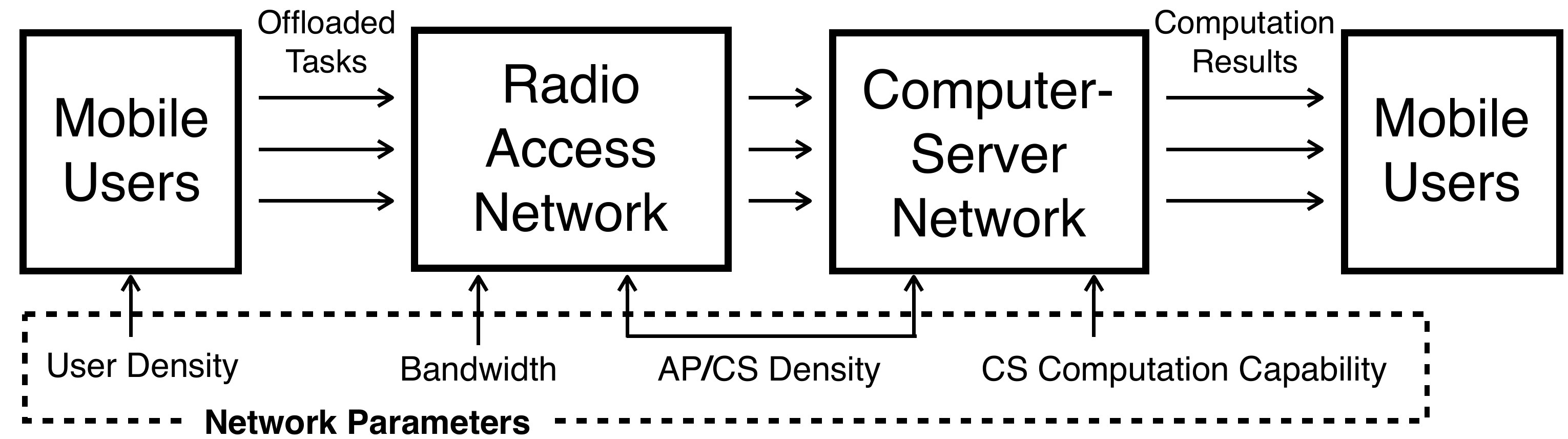}}
\vspace{-10pt}
\caption{The decomposition view of the MEC network.}\label{MEC_network}
\vspace{-30pt}
\end{figure}

\begin{figure}[t]
\centering
\includegraphics[width=6.5cm]{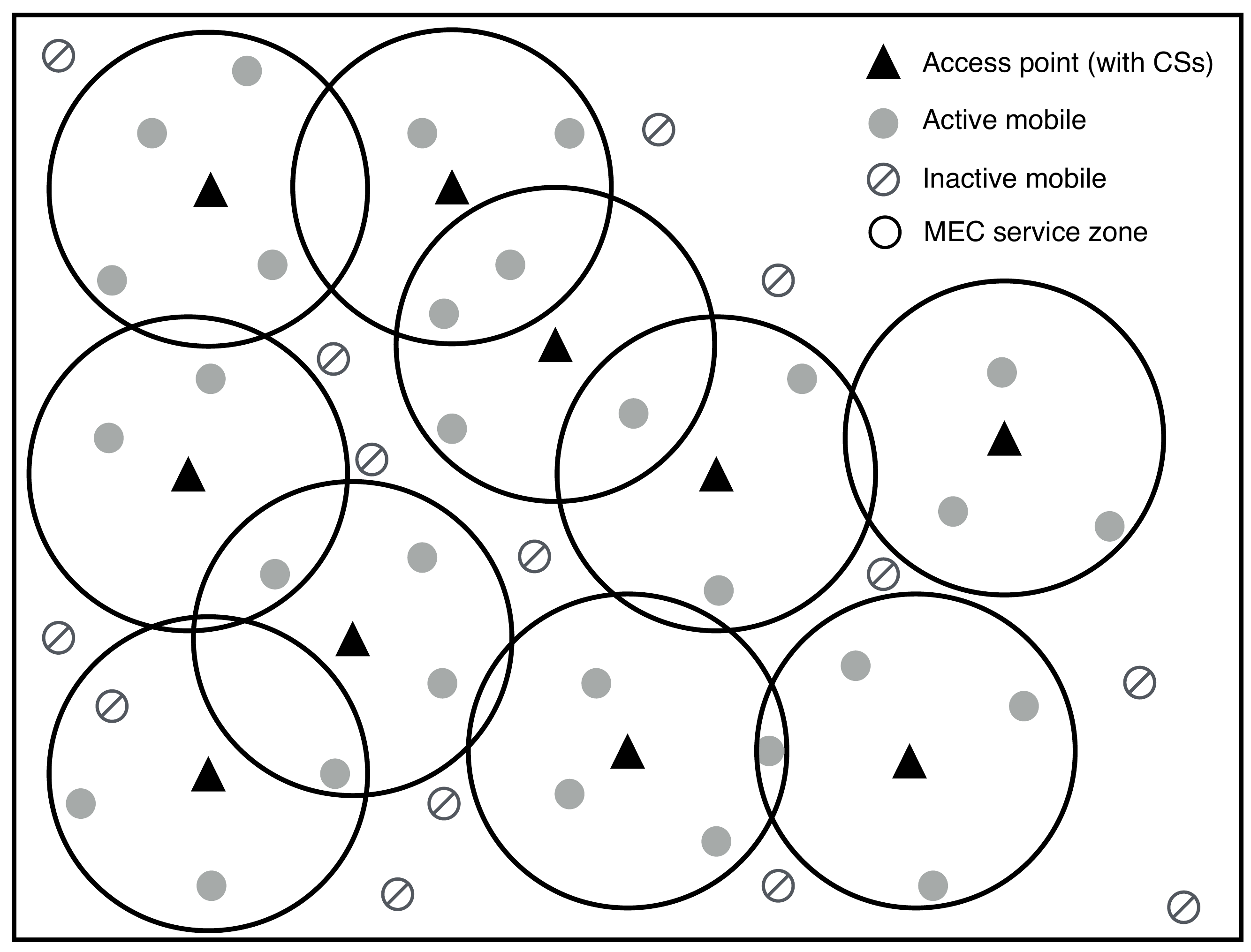}
\vspace{-10pt}
\caption{The spatial model of a MEC network.}\label{MEC_topology}
\vspace{-30pt}
\end{figure}

\vspace{-10pt}
\subsection{Modeling  Wireless Networks for Mobile Edge Computing }

In the past decade, stochastic geometry has  been established as a standard tool for modeling  and designing wireless networks, creating  an active research area \cite{elsawy2013stochastic}. A rich set of spatial  point processes such as \emph{Poisson point process} (PPP) and cluster processes have been used to model node locations in a wide range of wireless networks such as cellular networks \cite{Andrews:TractableApproachCoverageCellular:2010}, heterogeneous networks \cite{Dhillon:HeterNetModelAnalysis:2012}, and cognitive radio networks \cite{Hossain:2015Cognitive}. Based on  these network models and applying mathematical tools from stochastic geometry, the effects of most key  physical-layer techniques on network performance have been investigated  ranging from multi-antenna transmissions \cite{Vaze2012} to multi-cell cooperation \cite{Lin2014}. Recent advancements in the area can be found in numerous surveys  such as \cite{HaenggiAndrews:StochasticGeometryRandomGraphWirelessNetworks}. Most existing work in this area shares the same theme of how to cope with interference and hostility of wireless channels (e.g., path loss and fading) so as to ensure high coverage and link reliability for \emph{radio access  networks} (RAN) or distributed device-to-device networks. In contrast, the design of large-scale MEC networks in Fig.~\ref{MEC_system} has different objectives, all of which should jointly address two aspects of network performance, namely \emph{wireless communication and edge computing}.

Modeling a MEC network poses new challenges as its architecture  is  more complex than a traditional RAN and can be decomposed as a RAN cascaded with a {\emph{computer-server network}} (CSN) as illustrated in Fig. \ref{MEC_network}. 
The power of modeling MEC networks using stochastic geometry lies in allowing network performance to be described by a function of a relatively small  set of network parameters. To be specific, as shown in Fig. \ref{MEC_network}, the process of mobiles is parametrized by mobile density, the RAN by channel bandwidth and \emph{access-point} (AP) density, and the CSN by  CS density and CS computation capacity.  Besides the parameters, the performance of a MEC network is measured by numerous metrics. Like small-scale systems (see e.g., \cite{kao2017hermes} and \cite{liu2016delay}), the \emph{link-level} performance of the MEC network is measured by latency, which can be divided into latency for offloading in the RAN, called \emph{communication latency} (comm-latency) and latency for computing at CSs, called \emph{computation latency} (comp-latency). At the network level,  the coverage   of the RAN of a MEC network  is typically measured by   \emph{connectivity  probability} (also called coverage probability \cite{HaenggiAndrews:StochasticGeometryRandomGraphWirelessNetworks}), quantifying  the fraction of users having reliable links to APs.  A similar metric, called  \emph{stability probability}, can be defined for measuring the stability of  the CSN, quantifying the fraction of CSs  having finite comp-latency. There exist potentially complex relations between these four metrics that are regulated by the said network parameters. Existing results focusing solely on RAN (see e.g., \cite{HaenggiAndrews:StochasticGeometryRandomGraphWirelessNetworks})  are insufficient for quantifying these relations. Instead, it calls for developing a more sophisticated analytical  approach integrating theories of stochastic geometry, queueing, and parallel computing. 

Last, it is worth mentioning that comm-latency and comp-latency have been extensively studied in the literature mostly for point-to-point systems using queueing theory (see e.g., \cite{BertsekasBook:DataNetwk:92, barbarossa2014communicating}). However, studying such latency in large-scale networks is  much more challenging due to the existence of interference between randomly distributed nodes. As a result, there exist only limited results  on comm-latency in such networks \cite{martin2013local, gong2013mobiledelay, zhong2016staticpoisson}. In \cite{martin2013local},  the comm-latency given retransmission  is derived using stochastic geometry for the extreme cases with either static nodes or nodes having high mobility. The analysis is generalized in \cite{gong2013mobiledelay} for  finite mobility. 
Then the approach for comm-latency as proposed  in \cite{martin2013local} and \cite{gong2013mobiledelay} is further developed in \cite{zhong2016staticpoisson} to integrate stochastic geometry and queueing theory. Compared with these studies, the current work considers a different type of network, namely the MEC network, and  explores a different research direction, namely the tradeoff between comm-latency and comp-latency under constraints on the mentioned network-level performance metrics.

\vspace{-20pt}
\subsection{Contributions}
\vspace{-5pt}

{This work represents the first attempt on modeling a large-scale MEC network using stochastic geometry. The proposed model has  several features admitting tractable analysis of network latency performance.  
First, the locations of co-located pairs of CS and AP and the mobiles are distributed as two independent  homogeneous PPPs. 
Second, multiple access is enabled by spread spectrum \cite{Verdu1999}, which underpins the technology of  code-domain \emph{Non-Orthogonal Multiple Access} (NOMA) to be deployed in 5G systems for enabling massive access \cite{dai2015non}. Using the technology, interference is  suppressed by a parameter called spreading factor, denoted as $G$, at the cost of  data bandwidth reduction. 
Third, each mobile randomly generates a computation task in every time slot. 
Last, each CS computes multiple tasks simultaneously by parallel computing realized via  creating a number of \emph{virtual machines} (VMs), where the so called \emph{input/output} (I/O) interference in parallel computing is modelled \cite{bruneo2014stochastic}.

In this work, we propose an approach building on the spatial network model and the joint applications of tools from diversified areas including stochastic geometry, queueing, and parallel computing.  Though the network performance analysis relies on  well known tools, their applications are far more than straightforward. In fact, new challenges arise from the coupling of communication and edge computing in the MEC network. For example, the simple server model (with memoryless service time)  in the traditional queueing theory is now replaced  with a more complex MEC server model  featuring dynamic virtual machines and their I/O interference. As another example, the random computing-task arrivals are typically modelled as a single stochastic process  in conventional computing/queueing  systems but the current model has to  account for numerous network features ranging from  random node  distributions to  multiple access. The complex network model introduces new technical challenges that call for the development of a systematic framework for studying the MEC network performance and deployment,  which forms the theme of this work.
The  main contributions are summarized~below.
\begin{itemize}
\item {\bf Modeling a MEC network using stochastic geometry}: As mentioned, this work presents a novel model of a large-scale MEC network constructed using stochastic geometry. Given the complexity of the network, the contribution  in network modeling  lies in proposing a  model that is not only sufficiently practical but at the same time allows a tractable approach of analyzing network latency performance,  by integrating stochastic geometry, parallel computing, and queuing theory. The results and insights are summarized as follows.
\item {\bf Communication latency}: The expected comm-latency for an offloaded task, denoted as $\mathsf{T}_{\textrm{comm}}$,  is minimized under a constraint on the network connectivity probability. This is transformed into a constrained optimization problem of the spreading factor $G$. Solving the problem  yields the minimum $\mathsf{T}_{\textrm{comm}}$. The result shows that when mobiles are sparse, the full bandwidth should be allocated for data transmission so as to minimize $\mathsf{T}_{\textrm{comm}}$. However, when mobiles are dense, spread spectrum with large $G$ is needed  to mitigate interference for satisfying the  network-coverage constraint, which increases $\mathsf{T}_{\textrm{comm}}$. As a result, the minimum $\mathsf{T}_{\textrm{comm}}$ diminishes   inversely proportional to the channel bandwidth and as a power function of   the allowed fraction of disconnected users with a negative exponent, but grows sub-linearly with the expected number of mobiles per AP (or CS). In addition, $\mathsf{T}_{\textrm{comm}}$ is a monotone increasing function of the task-generation probability per slot that saturates   as the probability approaches one.
\item {\bf Analysis of RAN offloading throughput}: The RAN throughput, which determines  the load of the CSN (see Fig. \ref{MEC_network}), can be measured by 
the expected task-arrival rate at a typical AP (or CS). The rate is shown to be a \emph{quasi-concave} function of  the expected number of mobiles per AP, which  first increases  and then decreases as the ratio grows. In other words, the expected task-arrival rate is low in both   sparse  and dense networks. The maximum rate is proportional to  the bandwidth.  
\item {\bf Computation latency Analysis}:  
First, to maximize CS computing rates, it is shown that the dynamic number of VMs at each CS  should be no more than a derived number to avoid suffering rate loss due to their I/O interference. Then to ensure stable CSN, it is shown that the resultant maximum computing rate should be  larger than the task-arrival rate 
scaled by a factor larger than one, which  is determined by the allowed fraction of unstable CSs. Based on the result for parallel computing, tools from stochastic geometry and M/M/m queues are applied to derive bounds on the expected comp-latency for an offloaded task, denoted as $\mathsf{T}_{\textrm{comp}}$. The bounds show that the latency is \emph{inversely proportional} to the maximum computing rate  and \emph{linearly proportional} to the total task-arrival rate at the typical CS (or AP). Consequently, $\mathsf{T}_{\textrm{comp}}$ is a quasi-concave function of the expected number of users per CS (or AP) while $\mathsf{T}_{\textrm{comm}}$ is a monotone increasing function.
\item {\bf Network provisioning and planning}: Combining the above results suggest the following guidelines for network provisioning and planning. Given a mobile density, the AP density should be chosen for maximizing the RAN offloading throughput under the network-coverage constraint. Then sufficient bandwidth should be provisioned to simultaneously achieve the targeted   comm-latency for offloading a task. Last, given the mobile and RAN parameters, the CS computation capacities  are planned to achieve the targeted comp-latency for a offloaded task as well as enforcing the network-stability constraint.  The derived analytical results simplify  the calculation in the specific planning process. 
\end{itemize}


\vspace{-10pt}
\section{Modeling MEC Networks}\label{Network_model}
In this section, a mathematical model of the  MEC network as illustrated  in Fig.~\ref{MEC_system} is presented. 

\vspace{-15pt}
\subsection{Network Spatial Model}\label{System_Model}
\vspace{-5pt}
APs (and thus their co-located CSs) are randomly distributed in the horizontal plane and are modelled as a  homogeneous PPP $\Omega = \{Y\}$ with density $\lambda_b$, where $Y \in \mathds{R}^{2}$ is the coordinate of the corresponding AP. Similarly, mobiles are modelled as another    homogeneous PPP $\Phi = \{X\}$ independent of $\Omega$ and  having the  density $\lambda_m$.

Define a \emph{MEC-service zone} for  each AP,  as a disk region centered at $Y$ and having a fixed radius $r_0$, denoted by $\mathcal{O}(Y, r_0)$, determined by the maximum uplink transmission power of each mobile (see Fig.~\ref{MEC_topology}).
A mobile can access a AP for computing if it is covered by the MEC-service zone of the AP. It is possible that a mobile is within the service ranges of more than one AP. In this case, the mobile randomly selects a single AP to receive the MEC service. As illustrated in  Fig.~\ref{MEC_topology},  combining the randomly located MEC-service zones, $\cup_{Y \in \Omega} \mathcal{O}(Y, r_0)$,  forms a \emph{coverage process}. Covered mobiles are referred to as  \emph{active} ones and others \emph{inactive} since they remain silent. To achieve close-to-full network coverage, let the fraction of inactive mobiles be no more than  a small positive number $\delta$. Then the radius of MEC-service zones, $r_0$, should be set as $r_0 = \sqrt{\frac{\ln \frac{1}{\delta}}{\pi\lambda_b }}$ \cite{HaenggiAndrews:StochasticGeometryRandomGraphWirelessNetworks}. Given $r_0$, the number of mobiles  covered by an arbitrary MEC service zone follows a   Poisson distribution with mean $\lambda_m \pi r_0^2$. Consider a typical AP located at the origin. Let $X_0$ denote a  typical mobile located in the typical MEC service zone $\mathcal{O}(o, r_0)$. Without loss of generality, the network performance analysis focuses on the typical mobile.

\vspace{-15pt}
\subsection{Model of Mobile Task Generation}\label{Subsection:MobileTaskArrivalDeparture}
\vspace{-5pt}
Time is divided into slots having a unit  duration. Consider an arbitrary mobile.
A computation task is randomly generated in each slot with probability $p$, referred to as the  \emph{task-generation  rate}\setlength{\skip\footins}{0.1cm}\footnote{{The random task generation is an abstracted model allowing tractable analysis, and it is widely used in the literature in the same vein. 
The statistics of task generation can be empirically measured by counting the number of user service requests, which is shown in \cite{He2014} and \cite{Chang2008} to be bursty and periodical. It is interesting to use a more general task generation model, which is outside the scope
of current work. }}. The generated tasks are those favorable offloading in terms of energy efficiency such that offloading can save more energy than the local computing. The analysis on the offloading favorable condition will be given in the sequel. Task generations over two different slots are assumed to be independent.  The mobile has a unit buffer to store at most a single task for offloading. A newly generated task is sent for offloading when the buffer is empty or otherwise computed locally. This avoids significant queueing delay that is unacceptable in the considered case of latency-sensitive mobile computation. For simplicity, offloading each   task is assumed to require transmission of a fixed amount data. The transmission of a single task occupies a single \emph{frame} lasting  $L$ slots. The mobile checks whether the buffer is empty at the end of every $L$ slots and transmits a stored task to a serving AP.  Define the  \emph{task-offloading probability} as the probability that  the mobile's buffer  is occupied, denoted as $p_L$. Equivalently, $p_L$ gives the probability that   at least one task is generated within one frame:
\begin{align}\label{Eq:QueueOccupiedProb}
p_L=1-(1-p)^L.
\end{align}
Thereby, the task-departure process at a mobile  follows a Bernoulli process  with parameter $p_L$ provided the radio link is reliable (see discussion in the sequel).

\vspace{-15pt}
\subsection{Radio Access Model}
\vspace{-5pt}

Consider  an uplink channel with the fixed  bandwidth of  $B$ Hz. The channel is shared by all mobiles for  transmitting data containing offloaded tasks  to their serving APs. The CDMA (or code-domain NOMA) is applied to enable multiple access. For CDMA based on the \emph{spread-spectrum technology},  each mobile \emph{spreads} every transmitted symbol by multiplying it with    a  \emph{pseudo-random} (PN) sequence of \emph{chips} ($1$s and $-1$s), which is generated  at a much higher rate than the symbols and thereby spreads the signal spectrum \cite{Verdu1999}. The multiple access of mobiles is enabled by assigning unique PN sequences to individual users. A receiver then retrieves the signal sent by  the desired transmitter by multiplying the   multiuser signal with the corresponding PN sequence. The operation suppresses inference and \emph{de-spreads} the signal spectrum to yield symbols.  Let $G$ denote the \emph{spreading factor} defined as the ratio between the  chip rate and   symbol rate, which is equivalent to the number of available PN sequences.  The cross-correlation of PN sequences  is proportional to $\frac{1}{G}$ and approaches to zero as $G$ increases. As a result, the interference power is reduced by the factor of $G$ \cite{Verdu1999}.\setlength{\skip\footins}{0.1cm}\footnote{For the special case of synchronous multiuser transmissions, orthogonal sequences (e.g., Hadamard sequences) can be used instead of PN sequences to achieve orthogonal access \cite{Verdu1999}. However, the maximum number of simultaneous users is $G$, making the design unsuitable for massive access.} On the other hand, the price for spread spectrum is that  the bandwidth available to individual mobiles is reduced by $G$, namely  $\frac{B}{G}$. 
\vspace{-15pt}
\begin{remark}[CDMA vs. OFDMA] \emph{While CDMA is expected to enable non-orthogonal access in next-generation systems, orthogonal frequency division multiple access (OFDMA) has been widely deployed in existing system. However, OFDMA limits the number of simultaneous users to be no more than the number of orthogonal sub-channels. Compared with OFDMA, CDMA separates different users by PN sequences. The number of possible PN sequences can be up to $2^G -1$ with $G$ being the spreading factor (sequence length). In theory, an equal number of simultaneous users can be supported by CDMA that can be potentially much larger than that by OFDMA. Allowing non-orthogonality via CDMA provides a graceful tradeoff between the system-performance degradation and the number of simultaneous users, facilitating massive access in 5G. The current analysis of comm-latency can be straightforwardly extended to OFDMA by removing interference between scheduled users. For unscheduled users, comm-latency should include scheduling delay and the corresponding analysis is standard (see e.g., \cite{zhong2016}).}
\end{remark}
\vspace{-10pt}
Uplink channels are characterized by  path-loss  and small-scale Rayleigh fading. Assuming  transmission  by  a mobile with the fixed power $\eta$, the received signal power at the AP is given by $\eta g_X |Y - X|^{-\alpha}$, where $\alpha$ is the path-loss exponent, the $\exp(1)$ \emph{random variable} (RV) $g_X$ represents Rayleigh fading  and $|X-Y|$ denotes the Euclidian distance between $X$ and $Y$. Based on the channel model, the power of  interference at the  typical AP $Y_0$, denoted by~$I$, can be  derived as follows. Among potential interferers for the typical AP, the fraction of $\delta$ is  outside MEC-service zones. 
Given random task generation discussed earlier, each interferer transmits with probability $p_L$. Consequently, the active interferers form a PPP given by   $\tilde{\Phi}$ 
 with density $(1-\delta)p_L\lambda_m$ resulting  from thinning $\Phi$. 
It follows that the interference power $I$ can be written as $I = \frac{1}{G}\sum_{X \in \tilde{\Phi}}  \eta g_X |X|^{-\alpha}$, where the factor $\frac{1}{G}$ is due to the spread spectrum. Consider an interference-limited radio-access network where channel  noise is negligible.
The received $\SIR$ of the typical mobile is thus given as
\begin{align}\label{Eq:SIR}
\SIR_0 = \frac{g_{X_0} |X_0|^{-\alpha}}{\frac{1}{G}\sum_{X \in \tilde{\Phi}} \eta g_X |X|^{-\alpha}}.
\end{align}
The condition for successful offloading is that $\SIR$ exceeds a fixed threshold $\theta$ depending on the coding rate. Specifically, given $\theta$, the spectrum efficiency is $\log_2(1+\theta)$ (bits/sec/Hz) \cite{HaenggiAndrews:StochasticGeometryRandomGraphWirelessNetworks}. It follows that to transmit a task having a size of $\ell$ bits within a frame, the frame length $L$ should satisfy $L =\frac{G\ell}{B \cdot t_0 \cdot \log_2(1+\theta)}$ (in slots) where $t_0$ is the length of a slot (in sec). Define the minimum time for transmitting a task using the full bandwidth $B$ as  $\mathsf{T}_{\min} = \frac{\ell}{B \cdot t_0\cdot \log_2(1+\theta)}$ for ease of notation, giving $L = G\mathsf{T}_{\min}$.

\vspace{-5pt}
\begin{assumption}[Slow Fading]\label{ass.2}
\emph{We assume that channels  vary  at a much slower  time scale than that for mobile computation. To be specific, the mobile  locations and channel coefficients $\{g_X\}$ remain fixed  in the considered time window of computation offloading.}
\end{assumption}
\vspace{-25pt}
\noindent {
\begin{remark}[Fast Fading] \emph{In the presence of sufficiently high mobility, the channel variation can be faster than edge computation, resulting in fast fading. In this case, a mobile facing an unfavorable
channel can rely on retransmission to exploit the channel variation for reliable offloading. Nevertheless, this results in retransmission delay and thereby increases comm-latency. It is straightforward to analyze the extra latency in a large-scale network by applying an existing method (see e.g., \cite{martin2013local})}.
\end{remark}}
\vspace{-10pt}
By Assumption \ref{ass.2},   mobiles' $\SIR$s  remain constant and thereby mobiles can be separated into \emph{connected} and \emph{disconnected} mobiles.  To be specific, a mobile is connected to an AP if the corresponding SIR is above the threshold $\theta$ or otherwise disconnected.

\begin{figure}[t]
\centering
\begin{minipage}{0.495\textwidth}
\centering
\vspace{-5pt}
\subfigure[Synchronous offloading.]{\includegraphics[width=6 cm]{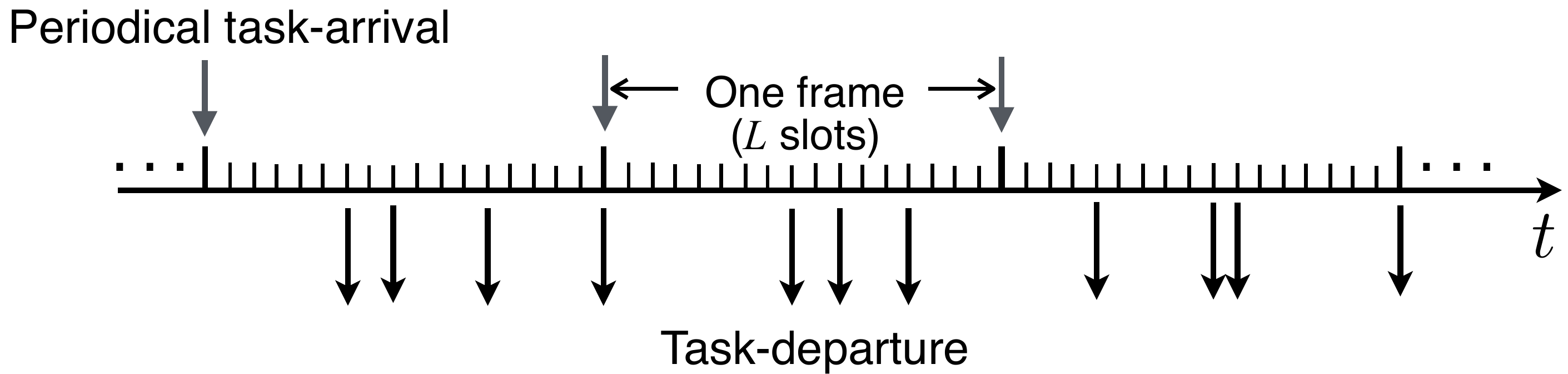}}\vspace{-5pt}\\
\subfigure[Asynchronous offloading.]{\includegraphics[width=6 cm]{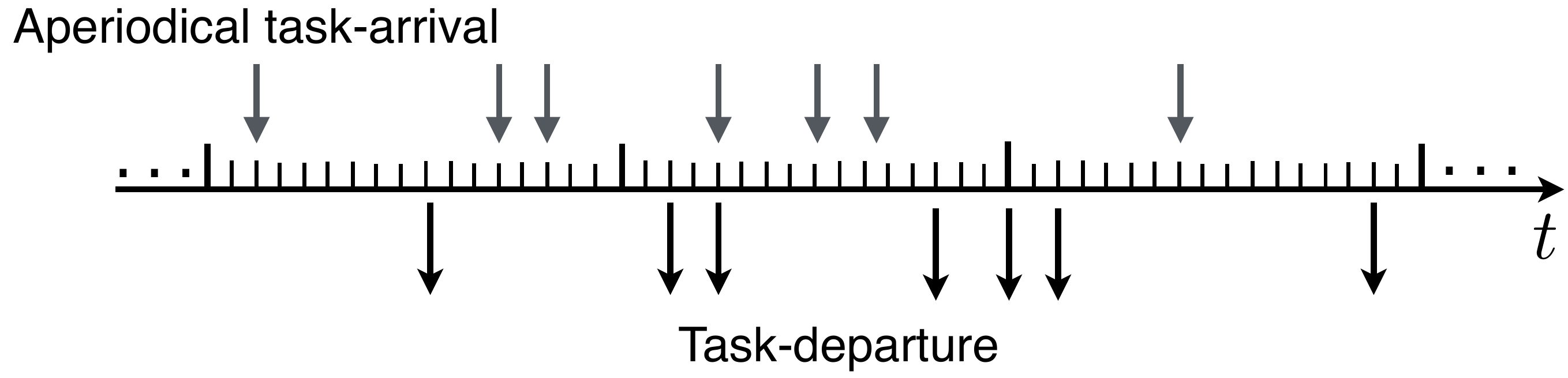}}
\vspace{-10pt}
\caption{Task arrival \& departure of two offloading modes. }
\label{fig_asyncsyncoffloading}
\end{minipage}
\begin{minipage}{0.495\textwidth}
\centering
\includegraphics[width=6 cm]{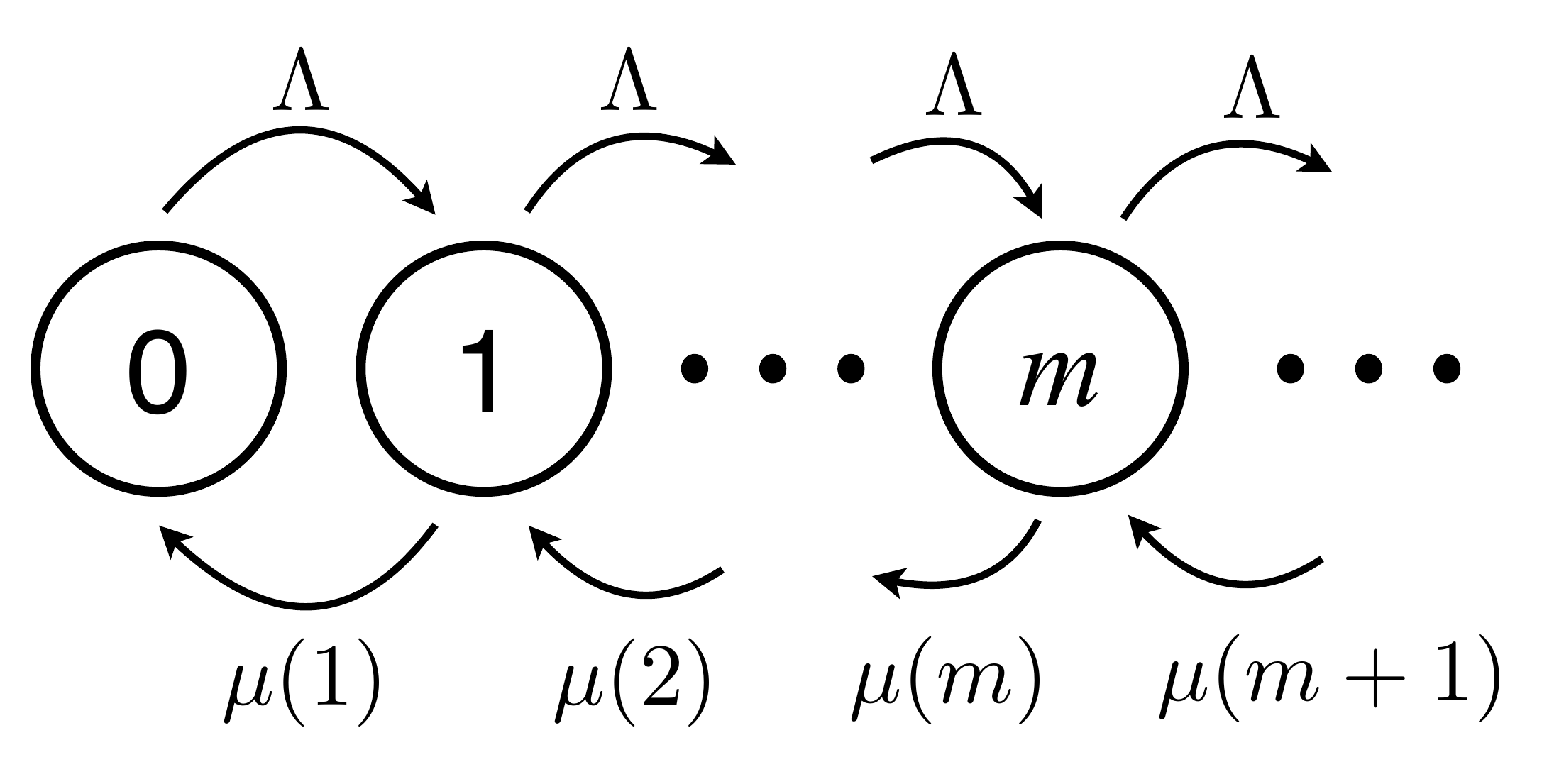}
\vspace{-5pt}
\caption{Markov chain modeling the tasks queueing for computation at the typical CS where $\Lambda$ is the arrival rate and $\mu(m)$ is the computation rate given $m$ waiting tasks.  }\label{mmc_fig}
\end{minipage}
\vspace{-30pt}
\end{figure}

We consider both \emph{synchronous} and \emph{asynchronous} multiuser transmissions defined in existing wireless standards such as 3GPP LTE. For synchronous transmissions, the frame boundaries of different users are aligned so as to facilitate protocols such as control signaling and channel feedback. Synchronization incurs network overhead for implementing a common clock as well as increases latency. For asynchronous transmissions, the said constraint on frame boundaries is not applied and thus  the transmission of each mobile is independent of those of others. The  transmissions modes lead to different task-arrival models for CSs. Specifically, given synchronous transmissions, the offloaded tasks arrive at a CS in batches and periodically as illustrated in Fig. \ref{fig_asyncsyncoffloading}(a).
The number of arrival tasks in each batch  is random depending on the number of connected mobiles in the same  MEC-service zone. On the other hand, given asynchronous transmissions, the offloaded tasks arrive at an AP at different time instants as illustrated in Fig. \ref{fig_asyncsyncoffloading}(b).

\vspace{-15pt}
\subsection{Edge-Computing Model}\label{Subsection:EdgeComputingModel}

\subsubsection{Parallel-Computing Model}\label{Subsubsection:EdgeComputingModel_ParallelComp}
 Upon their arrivals at APs, tasks are assumed to be delivered to CSs without any delay and queue at the CS buffer for computation on the first-come-first-served basis. Moreover, each CS is assumed to be provisioned with large storage modelled as a buffer with infinite capacity. At each CS, parallel computing of multiple tasks is implemented by creating \emph{virtual machines} (VM) on the  same \emph{physical machine} (PM) \cite{bruneo2014stochastic}. VMs are created asynchronously such that a VM can be added or removed at any time instant. It is well known in the literature that simultaneous VMs interfere with each other due to their sharing common computation resources in the  PM e.g.,  CPU, memory, buses for I/O. The effect is called \emph{I/O interference} that reduces the computation speeds of VMs.    The model of I/O interference as proposed in  \cite{bruneo2014stochastic} is adopted where the expected computation time for  a single  task\footnote{{The latency caused by creating and releasing VMs is not explicitly considered. It
is assumed to be part of computation time.}}, denoted by $T_c$, is  a function of the number of VMs,  $m$:
\begin{equation}
T_c(m) = T_0 (1+d)^{m-1}, \label{T_c}
\end{equation}
where $T_0$ is the expected computation time of a task in the case of a single VM ($m=1$) and $d$ is the degradation factor due to I/O interference between VMs. One can observe that $T_c$ is a monotone increasing function of $d$.  For tractability, we assume that the computation time  for a task is an $\exp(T_c)$ RV following the common assumption  in queueing theory \cite{bruneo2014stochastic}.

\subsubsection{CS Queuing  Model} The general approach of analyzing comp-latency relies on the interplay between parallel-computing and queueing theories. In particular, for the case of asynchronous offloading, the task arrival at the typical AP is approximated as a Poisson process for the following reasons. Due to the lack of synchronization between mobiles, the time instants of tasks arrivals are approximately uniform in time. Furthermore, at different time instants, tasks are generated following i.i.d. Bernoulli distributions based on the model in Section \ref{Subsection:MobileTaskArrivalDeparture}. {It is well known that the superposition of independent arrival process behaves like a Poisson process~\cite{sriram1986characterizing}.}

\vspace{-10pt}
\begin{assumption}\label{Assumption:PoissonArrival}\emph{For  the case of asynchronous offloading, given $N$ connected mobiles and the spreading factor $G$, the task arrivals at the typical AP are approximated as a Poisson process with the arrival rate of $\Lambda(N, G)=  \frac{N p_L }{L} =  \frac{N p_L }{G \mathsf{T}_{\min}}$.
}
\end{assumption}
\vspace{-10pt}
The Poisson approximation is   shown  by  simulation to be accurate in Appendix III. Given the Poisson arrival process and exponentially distributed computation time, the random number of tasks queueing at  the typical CS can be modelled as a continuous-time Markov chain as illustrated in Fig. \ref{mmc_fig} \cite{BertsekasBook:DataNetwk:92}. In the Markov chain, $\Lambda$ denotes the task-arrival rate in  Assumption \ref{Assumption:PoissonArrival} and  $\mu(k)$ denotes the CS-computation rate (task/slot) given $k$ tasks in the CS. The CS-computation rate is maximized in the sequel by optimizing the number of VMs based on the queue length.

Last, the result-downloading phase is not considered for brevity. First, the corresponding latency analysis is similar to  that for the offloading phase. Second,  the latency for downloading   is  negligible compared with those for offloading. The reasons are that computation results typically have small sizes compared with offloaded tasks and furthermore   downlink  transmission rates are typically much higher than uplink  rates. 

\vspace{-10pt}
\subsection{Performance Metrics}
The network performance is measured by two metrics: \emph{comm-latency} and \emph{comp-latency}. The definitions of metrics build on the  design constraints for ensuring  network connectivity and stability defined as follows.
\vspace{-10pt}
\begin{definition}[Network Coverage  Constraint]\label{Def:epsilonStability}\emph{The RAN in Fig. \ref{MEC_network} is designed to be \emph{$\epsilon$-connected}, namely that the portion of mobiles  is no less   than $(1-\epsilon)$, where $0< \epsilon \ll 1$.}
\end{definition}
\vspace{-10pt}
\noindent The fraction of connected mobiles is equivalent to the \emph{success   probability}, a metric widely used for studying the performance  of random wireless networks \cite{HaenggiAndrews:StochasticGeometryRandomGraphWirelessNetworks}. For the MEC network, the success probability is renamed as \emph{connectivity probability} and defined for the typical mobile as the following function of the spreading factor $G$:
\begin{align}\label{ps_def}
\mathsf{p}_c(G) = \mathrm{Pr}\l( \SIR_0 \geq \theta \r),
\end{align}
where $\SIR_0$ is given in \eqref{Eq:SIR}. Then the network coverage constraint can be written as $\mathsf{p}_c(G) \geq (1 - \epsilon)$. Under the connectivity constraint, most mobiles are connected to APs. Then the  comm-latency, denoted as $\mathsf{T}_{\textrm{comm}}$,  is defined as the \emph{expected} duration required for a connected mobile to offload a task to the connected AP successfully. The latency  includes both waiting time at the mobile's buffer and and the transmission time.

Next, consider the computation load of the typical  AP. Since the number of mobiles connected to the AP is a RV, there exists non-zero probability that the AP is overloaded, resulting in infinite queueing delay. In this case, the connected mobiles are referred to as being \emph{unstable}. To ensure most mobiles are stable, the following constraint is applied on  the network design.
\vspace{-10pt}
\begin{definition}[Network Stability Constraint]\label{Def:varepsilonStability}\emph{The CSN in Fig. \ref{MEC_network} is designed  to be \emph{$\rho$-stable}, namely that  the fraction  of stable CSs is no less  than $(1-\rho)$, where $0< \rho\ll 1$. }
\end{definition}
\vspace{-10pt}
\noindent The fraction  $\rho$ is equivalent to the probability that the typical CS is stable, denoted as $\mathsf{p}_s$.  Under the stability  constraint, most connected mobiles are stable. Then the comp-latency, denoted by $\mathsf{T}_{\textrm{comp}}$, is defined for the typical connected mobile as the \emph{expected} duration from the instant when  an offloaded task arrives at  the serving   CS until the instant when the computation of the task is completed, which includes both queueing delay and actual computation time.

Last, given the above definitions, the network is referred to as being \emph{communication-limited} (comm-limited) if  $\mathsf{T}_{\textrm{comm}} \gg \mathsf{T}_{\textrm{comp}}$ and \emph{computation-limited} (comp-limited) if $\mathsf{T}_{\textrm{comm}} \ll \mathsf{T}_{\textrm{comp}}$.

\vspace{-10pt}
\section{Communication Latency Analysis}\label{Section:CommLat}

In this section, the comm-latency defined in the preceding section is analyzed building on results from the literature of network modeling using stochastic geometry. Then the latency is minimized by optimizing the spreading factor for CDMA, which regulates the tradeoff between the transmission rates of connected mobiles and network-connectivity performance.

\vspace{-20pt}
\subsection{Feasible Range of Spreading Factor}
\vspace{-5pt}
As mentioned, the spreading factor $G$ is a key network parameter regulating the tradeoff between network coverage and comm-latency. To facilitate subsequent analysis, under the network constraint in Definition \ref{Def:epsilonStability}, the feasible range of  $G$ is derived as follows. The result is useful for minimizing the comm-latency in the next sub-section. To this end, consider the connectivity probability defined in \eqref{ps_def}. Using a similar approach as the well-known one   for deriving network success probability using stochastic geometry (see e.g.,  \cite{Andrews:TractableApproachCoverageCellular:2010}),  we obtain the following result with the  proof omitted for brevity.
\vspace{-10pt}
\begin{lemma}[Connectivity Probability]\label{lemma:successOffProb}\emph{Given the spreading factor $G$, the connectivity probability of a typical mobile is given as
\begin{align}
\mathsf{p}_c(G) = \frac{1 - \exp\l(- \xi(G)\r)}{\xi(G)}, \label{ps}
\end{align}
where $\xi(G)$ is defined as
\begin{align}
\xi(G)=\frac{2(1-\delta)\l( 1 - (1-p)^{G \mathsf{T}_{\min}} \r)\ln \delta^{-1}}{\alpha} \mathcal{B}( \alpha ) \l( \frac{\lambda_m}{\lambda_b}\r)\l( \frac{\theta}{G} \r)^{\frac{2}{\alpha}},\label{xi}
\end{align}
and $\mathcal{B}( \alpha ) \triangleq \int_{0}^{1}\kappa^{\frac{2}{\alpha}-1} (1-\kappa)^{-\frac{2}{\alpha}}\mathrm{d} \kappa$ denotes the  Beta function.
}
\end{lemma}
\vspace{-10pt}
Recall that the network coverage constraint in Definition \ref{Def:epsilonStability} requires that $\mathsf{p}_c(G)\geq (1-\epsilon)$. Note that  $G$ is an important system parameter affecting both the transmission rates and the connectivity probability as elaborated in the following remark.
\vspace{-10pt}
\begin{remark}[Transmission Rates vs. Connectivity]\label{Remark:TRvsConnect}\emph{The  spreading factor $G$ of CDMA controls the tradeoff between mobile transmission rates and network connectivity probability. On one hand, increasing $G$  reduces the bandwidth, $\frac{B}{G}$, available to each mobile, thereby reducing  the transmission rate and increasing comm-latency. As the result, given longer frames with the task-generation rate being fixed, more mobiles are likely to have tasks for offloading at the beginning of each frame, increasing the density of interferers. On the other hand, growing $G$  suppresses interference power by the factor $G$ via spread spectrum. As a result, the   connectivity probability grows. Given the two opposite effects, one should expect that in the case of a stringent connectivity constraint, either small or large value for $G$ is preferred but no the moderate ones.
}
\end{remark}
\vspace{-10pt}

Next, the effects of the spreading factor as discussed in Remark \ref{Remark:TRvsConnect} are quantified by deriving the feasible range of $G$ under the connectivity constraint. Define  the Lambert function, $W(x)$, as the solution for the equation $W(x)e^{W(x)}=x$. Then using the result in Lemma \ref{lemma:successOffProb}, the coverage constraint $\mathsf{p}_c(G) \geq (1 - \epsilon)$  is equivalent to $\xi(G) \leq \mathcal{F}(\epsilon)$ with the function $\mathcal{F}(\epsilon)$ defined as
\begin{align}\label{xi_G_condition}
\mathcal{F}(\epsilon)= W\l( -\frac{e^{-\frac{1}{1-\epsilon}}}{1 - \epsilon} \r) + \frac{1}{1- \epsilon}.
\end{align}
Notice that $\lim_{\epsilon \rightarrow 0}\frac{d}{d\epsilon }W\l( -\frac{e^{-\frac{1}{1-\epsilon}}}{1 - \epsilon} \r) = 1$.
Moreover, $W\l( -\frac{e^{-\frac{1}{1-\epsilon}}}{1 - \epsilon} \r) = -1$ at $\epsilon = 0$. It follows that from these two results that $\mathcal{F}(\epsilon)$ can be approximated as 
\begin{equation}\label{Eq:F:Approx}
\mathcal{F}(\epsilon) \approx 2\epsilon, \qquad \epsilon \ll 1. 
\end{equation}

In addition, $\xi(G)$ is maximized at the point of $G=g_0$ of which the existence and uniqueness are proved in Lemma~\ref{lemma:exitsUniG}.
If $\xi(g_0)\leq \mathcal{F}(\epsilon)$, it is straightforward that any $G$ satisfies the condition of \eqref{xi_G_condition}.
Otherwise, the feasible range of $G$ satisfying the connectivity is provided in Proposition~\ref{lemma:feasibleG}.

\vspace{-10pt}
\begin{lemma}[Properties of  $\xi(G)$]\label{lemma:exitsUniG}\emph{The function $\xi(G)$  in \eqref{xi} attains its maximum at $G = g_0$ with
\begin{align}\label{Eq:g0}
g_0=\frac{\alpha W\l( -\frac{2}{\alpha}e^{-\frac{2}{\alpha}} \r) + 2}{{\alpha \mathsf{T}_{\min}} \ln(1-p)}.
\end{align}
Moreover, $\xi(G)$ is monotone increasing in the range $[-\infty, g_0]$ and monotone decreasing in the range $[g_0, \infty]$. }
\end{lemma}
\vspace{-10pt}
\begin{proof}
See Appendix \ref{proof:xiG}.
\end{proof}
\vspace{-10pt}
\begin{proposition}[Feasible Range of Spreading Factor]\label{lemma:feasibleG}\emph{Under the network connectivity constraint, the feasible range of $G$ is $G \geq 1$ if $\xi(g_0)\leq \mathcal{F}(\epsilon)$, where   $g_0$ is given in \eqref{Eq:g0}. If $\xi(g_0) > \mathcal{F}(\epsilon)$, the feasible range of $G$ is $\mathcal{S}=\mathcal{S}_1\bigcup \mathcal{S}_2$ where
\begin{equation}\label{Eq:S}
\mathcal{S}_1=\{G\in \mathds{Z}^+|1\leq G\leq g_a\},  \qquad \mathcal{S}_2=\{G\in \mathds{Z}^+|  G\geq g_b \},
\end{equation}
where $g_a$ and $g_b$ are the two roots of the equation $\xi(G) = \mathcal{F}(\epsilon)$. }
\end{proposition}
\vspace{-10pt}
Based  on Lemma \ref{lemma:exitsUniG}, the function $\xi(G)$ is monotone increasing over $\mathcal{S}_1$ but monotone decreasing over $\mathcal{S}_2$. In addition, if $g_a < 1$, $\mathcal{S}_1$ is empty and the feasibility range of $G$ reduces to $\mathcal{S}_2$.
\vspace{-15pt}
\subsection{Communication Latency}
\vspace{-5pt}
Recall that the comm-latency of connected mobiles $\mathsf{T}_{\textrm{comm}}$ comprises the expected waiting time for offloaded tasks at mobiles, denoted as $\mathsf{T}_{\textrm{comm}}^{(a)}$, and transmission delay, denoted as $\mathsf{T}_{\textrm{comm}}^{(b)}$. Consider the expected  waiting time. Recalling that  the offloading protocol  in  Section~\ref{Subsection:MobileTaskArrivalDeparture},   the first  task arrival  during $L$ slots is delivered to the offloading buffer  and the subsequent tasks  are forwarded to the local computation unit. Let $K$ denote the slot index when an offloaded  task arrives at the offloading buffer.  It follows that  the probability distribution of  $K$ follows a conditional geometric distribution, i.e., $\Pr(K=k)=\frac{p (1-p)^{k-1}}{1-(1-p)^L}$,
where $k = 1, 2, \cdots, L$ and the normalization  term $1-(1-p)^L$ gives  the probability that  at least one task arrives during a single  frame.
Thereby, the expected  waiting time is given as
\begin{align}\label{Eq:Tcompa}
\mathsf{T}_{\textrm{comm}}^{(a)} = \sum_{k=1}^L(L-k) \frac{p(1-p)^{k-1}}{1-(1-p)^L}
&= \frac{L}{1-(1-p)^L}-\frac{1}{p}.
\end{align}
Next, consider the transmission time for a single task in a frame that spans $L$ slots. Recall that $L=G\mathsf{T}_{\min}$ where $\mathsf{T}_{\min}$ is the minimum time for transmitting a task as defined earlier.  Combining $\mathsf{T}_{\textrm{comm}}^{(b)} = G\mathsf{T}_{\min}$ and $\mathsf{T}_{\textrm{comm}}^{(a)}$ in \eqref{Eq:Tcompa} gives the following result.
\vspace{-10pt}
\begin{lemma}[Comm-Latency]\label{the:retrans_latency}\emph{Given the spreading factor $G$,
the  comm-latency of the typical mobile $\mathsf{T}_{\textrm{comm}} $  (in slot) is given as
\begin{equation}
\mathsf{T}_{\textrm{comm}}(G)= G{\mathsf{T}_{\min}}+  \frac{G \mathsf{T}_{\min}}{1 - (1-p)^{G \mathsf{T}_{\min}}}  - \frac{1}{p}, \label{Eq:CommunLatency:a}
\end{equation}
where $\mathsf{T}_{\min}$ is the minimum time for transmitting a task using full bandwidth. }
\end{lemma}
\vspace{-10pt}
Next, consider the minimization of the comm-latency over the spreading factor $G$.  Using \eqref{Eq:CommunLatency:a}, it is straightforward to show that the comm-latency $\mathsf{T}_{\textrm{comm}}(G)$ is a monotone  increasing function of  $G$. Therefore, minimizing  comm-latency is equivalent to minimizing $G$. It follows from Proposition~\ref{lemma:feasibleG} that the minimum of $G$, $G^*=\underset{G\in{\mathcal{S}}}{\min}  G$, is given as
\begin{equation}\label{Eq:MinG}
G^* = \l\{
\begin{aligned}
&g_b, &&\mathcal{S}_1 = \emptyset,\\
&1, && \text{otherwise}.
\end{aligned}
\r.
\end{equation}
Substituting  $G^*$ into \eqref{Eq:CommunLatency:a} gives the minimum comm-latency as shown in the following theorem.
\vspace{-30pt}
\begin{theorem}[Minimum Comm-Latency]\label{theo:optimalG}\emph{By optimizing the spreading factor $G$, the minimum comm-latency (in slot), denoted as $\mathsf{T}_{\textrm{comm}}^* $,  is given as follows.
\begin{enumerate}
\item If $\mathcal{S}_1$ in \eqref{Eq:S} is non-empty,
\begin{equation}
\mathsf{T}_{\textrm{comm}}^*
= \mathsf{T}_{\min}+  \frac{\mathsf{T}_{\min} }{1 - (1-p)^{\mathsf{T}_{\min}}}  - \frac{1}{p}, \label{Eq:CommunLatency_}
\end{equation}
where $\mathsf{T}_{\min} = \frac{\ell}{B \cdot t_0 \cdot \log_2(1+\theta)}$.
\item If $\mathcal{S}_1$ is empty,
\begin{equation}
\mathsf{T}_{\textrm{comm}}^*
= g_b\mathsf{T}_{\min}+  \frac{g_b \mathsf{T}_{\min} }{1 - (1-p)^{g_b \mathsf{T}_{\min}}}  - \frac{1}{p}, \label{Eq:CommunLatency}
\end{equation}
where  $g_b$ is  specified in Proposition \ref{lemma:feasibleG}.
\end{enumerate}
}
\end{theorem}
\vspace{-10pt}
Consider the second case in Theorem \ref{theo:optimalG}. The comm-latency $\mathsf{T}_{\textrm{comm}}^*$ can be approximated in closed-form if $g_b\mathsf{T}_{\min}$ is sufficiently large. For this case, $\l[1 - (1-p)^{g_b \mathsf{T}_{\min}}\r]\approx 1$ and thus the function $\xi(G)$ in \eqref{xi} can be approximated as
\begin{equation}
\xi(G) \approx \frac{2(1-\delta)\ln \delta^{-1}}{\alpha} \mathcal{B}( \alpha ) \l( \frac{\lambda_m}{\lambda_b}\r) \l(\frac{\theta}{G}\r)^{\frac{2}{\alpha}}.
\end{equation}
It follows from Theorem \ref{theo:optimalG} and \eqref{Eq:F:Approx} that if $\mathcal{S}_1$ is empty and $g_b\mathsf{T}_{\min}$ is large,
\begin{equation}
\mathsf{T}_{\textrm{comm}}^*
\approx 2 g_b\mathsf{T}_{\min}  - \frac{1}{p}, \label{Eq:Tcomm}
\end{equation}
where
\begin{align}\label{Eq:OptimalGApprox}
g_b\approx \l[ \frac{(1-\delta)\ln \delta^{-1}\mathcal{B}( \alpha ) }{\alpha\epsilon }\l( \frac{\lambda_m}{\lambda_b}\r)\r]^{\frac{\alpha}{2}}\theta.
\end{align}

\begin{remark}[Sparse Network vs. Dense Network]\label{remark:depOptTcomMobile}\emph{The first and second cases in Theorem \ref{theo:optimalG} correspond to sparse and dense networks, respectively, as measured by the mobile-to-AP density ratio $\lambda_m/\lambda_b$. In the first case ($\mathcal{S}_1\neq \emptyset$), the network is sufficiently sparse, namely the  ratio  $\lambda_m/\lambda_b$ is sufficiently small, such that the optimal spreading factor $G^*=1$  and the resultant comm-latency is independent of the ratio as shown in the theorem. In other words, for this case, it is optimal to allocate all bandwidth for increasing the transmission rate instead of reducing it for the purpose of suppressing interference to satisfy the  network connectivity constraint. In contrast, in the second case  ($\mathcal{S}_1= \emptyset$), the network is relatively dense and it is necessary to apply spread spectrum to reduce interference so as to  meet the connectivity requirement, corresponding to $G^* > 1$. As the result, the minimum comm-latency scales with the density ratio as $\mathsf{T}_{\textrm{comm}}^*\propto \l(\frac{\lambda_m}{\lambda_b}\r)^{\frac{\alpha}{2}}$ as one can observe from \eqref{Eq:OptimalGApprox}.}
\end{remark}

\vspace{-15pt}
\begin{remark}[Effects of Network Parameters]\label{remark:commScaLaw}\emph{Substituting $\mathsf{T}_{\min} = \frac{\ell}{B \cdot t_0\cdot \log_2(1+\theta)}$ into \eqref{Eq:Tcomm} gives that for a relatively dense network, the comm-latency scales as 
\begin{equation}
\boxed{
\mathsf{T}_{\textrm{comm}}^*
\propto \frac{\ell}{B}\l(\frac{\lambda_m}{\epsilon \lambda_b}\r)^{\frac{\alpha}{2}}  - \frac{1}{p}.  
}
\end{equation}
The scaling laws shows the effects of network parameters including the task size $\ell$, bandwidth $B$, mobile  density $\lambda_m$ and AP density $\lambda_b$, and the task-generation probability per slot $p$. 
}
\end{remark}

\vspace{-30pt}
\subsection{Task-Arrival Rates at APs/CSs} \label{Subsection:TaskArrivalRate}
The offloading throughput of the RAN represents the load of the CSN (see Fig. \ref{MEC_network}). The throughput  can be measured by the  expected task-arrival rate (in number of tasks per slot) at the typical AP (equivalently the typical CS). Its scaling law with  the expected number of mobiles per AP, $\lambda_m/\lambda_b$, is not straightforward due to several factors. To be specific,  the total bandwidth is fixed,  the spread factor grows nonlinearly with  $\lambda_m/\lambda_b$, and the   likelihood of task-generation probability per frame varies with  the frame length. To address this issue, the task arrivals at the typical AP are characterized as follows. 

Consider the case of asynchronous offloading. Based on the model in Section \ref{Subsection:MobileTaskArrivalDeparture}, the probability that a mobile generates a  task for offloading in each frame is 
\begin{equation}
p_L^*=1 - (1-p)^{L^*}, \label{Eq:TaskOffloadingProb}
\end{equation}
where  $L^*$ is the frame length given the optimal spreading  factor $G^*$ in  \eqref{Eq:MinG}. The expected task-offloading  rate (in number of tasks per slot) for the typical mobile, denoted as $\beta^*$, is given as $\beta^* = \frac{p_L^*}{L^*}$. Since $L^* = G^*\mathsf{T}_{\min}$, 
\begin{equation}
\beta^*=\frac{1 - (1-p)^{G^*\mathsf{T}_{\min}}}{G^*\mathsf{T}_{\min}}\label{Eq:TaskOffloadingRate}. 
\end{equation}
where $\beta^*=p_L^* \cdot G^*\mathsf{T}_{\min}$. Let $\bar{\Lambda}^*$ denote the expected task-arrival rate at the typical AP (or CS). Then $\bar{\Lambda}^* = \bar{N}\beta^*$ where $\bar{N}$ is the expected number of mobiles connected to the AP. Since $\bar{N}=(1-\delta)(1-\epsilon)\frac{\lambda_m}{\lambda_b}$, 
\begin{equation}\label{Eq:TaskArrivalRate}
\bar{\Lambda}^* = (1-\delta)(1-\epsilon)\frac{\lambda_m}{\lambda_b}\beta^*. 
\end{equation}

\begin{remark}[Effects of Network Parameters]\label{Re:Para:Comm}\emph{Using \eqref{Eq:MinG}, \eqref{Eq:OptimalGApprox} and \eqref{Eq:TaskOffloadingRate}, one can infer that 
\begin{equation}\label{Eq:QuasiConcaveLambda}
\bar{\Lambda}^* \propto \l\{
\begin{aligned}
& p\frac{\lambda_m}{\lambda_b}, && \frac{\lambda_m}{\lambda_b}\rightarrow 0,\\
&B\l(\frac{\lambda_m}{\lambda_b}\r)^{-\frac{\alpha}{2}+1}, && \frac{\lambda_m}{\lambda_b}\rightarrow \infty. 
\end{aligned}
\r.
\end{equation}
The first case corresponds to a sparse network whose performance is not limited by  bandwidth and interference. Then the expected task arrival-rate grows linearly   with the task-generation probability per slot, $p$,  and the expected number of mobiles per AP, $\lambda_m/\lambda_b$. For the second case, in a dense network that is bandwidth-and-interference limited, the rate grows linearly the bandwidth $B$, but decreases with $\lambda_m/\lambda_b$. The reason for the decrease is the bandwidth for offloading is reduced so that a larger spreading factor is available for suppressing interference to meet the network-coverage requirement. Consequently,  the load for the CSs is lighter for a dense (thus comm-limited) network, reducing comp-latency as shown in the sequel. 
}
\end{remark}
\vspace{-10pt}
Consider tasks arrivals for the case of synchronous offloading. Unlike the asynchronous counterpart with arrivals spread over each frame, the tasks from mobiles arrive the typical AP at the beginning of each frame. Thus, it is useful to characterize the expected number of task arrivals per frame, denoted as $\bar{A}^*$, which can be written as $\bar{A}^* = \bar{N} p_L^*$. It follows that 
\begin{align}
\bar{A}^* =(1-\delta)(1-\epsilon)\frac{\lambda_m}{\lambda_b} p_L^*.  \label{Eq:AverageTaskArrival}
\end{align}
\begin{remark}[Effects of Network Parameters]\emph{In a dense network ($\lambda_m/\lambda_b\rightarrow \infty$), it can be obtained  from \eqref{Eq:MinG}, \eqref{Eq:OptimalGApprox}, and \eqref{Eq:TaskOffloadingProb} that  $p_L^* \approx 1$. Then it follows from \eqref{Eq:AverageTaskArrival} that the expected number of tasks per frame increases linearly with the expected number of mobiles per AP, $\lambda_m/\lambda_b$.}
\end{remark}

\vspace{-20pt}
\section{Computation Latency Analysis: Asynchronous Offloading}\label{Section:CompLat_asy}

This section aims at analyzing the comp-latency of the asynchronous offloading  where task arrival and departure are randomly distributed over time.
Given the Markov model of Fig.~\ref{mmc_fig}, we derive the network stability condition in Definition \ref{Def:varepsilonStability} and bounds of the average comp-latency.

\vspace{-15pt}
\subsection{Optimal Control of  VMs}

On one hand, creating a large number of VMs at the typical CS can slow down its computation rate due to the mentioned I/O interference between VMs. On the other hand, too few VMs can lead to marginal gain from parallel computing. Therefore, the number of VMs should be optimally controlled based on the number of waiting tasks. To this end, let $\mu(m)$ denote the computation rate given  $m$ VMs. Given the computation model in \eqref{T_c},  it follows from $\mu(m)=m/T_c(m)$ that:
\begin{equation}
 \mu(m) = \frac{m}{T_0}(1+d)^{1-m}.
\end{equation}
By analyzing the derivative of  $\mu(m)$, one can find that the function is monotone increasing before reaching a global maximum and after that it is monotone decreasing. Thereby, the value of $m$ that maximizes $\mu(m)$, denoted as $m_{\max}$,  can be found with the integer constraint 
\begin{align}\label{Eq:mmax}
m_{\max} = \textrm{round}\l(\frac{1}{\ln(1+d)}\r),
\end{align}
where $\textrm{round}(x)$ rounds $x$ to the nearest integer. The said properties  of the function $\mu(m)$ and the derived $m_{\max}$ in \eqref{Eq:mmax} suggest the following optimal VM-control policy.
\vspace{-10pt}
\begin{proposition}[Optimal VM Control]\label{lemma:optimalM}\emph{To maximize the computation rate at the typical CS, the optimal VM-control policy is to create $m_{\max}$ VMs if there are a sufficient number of tasks for computation or otherwise create as many VMs as possible until the buffer is empty. Consequently,
the maximum computation rate, denoted as  $\mu^*(m)$,  given $m$  tasks at the CS (being computed or in the buffer) is
\begin{equation}\label{Eq:muopt}
\mu^*(m) = \l\{
\begin{aligned}
&\frac{m}{T_0}(1+d)^{1-m}, &&1\leq m \leq m_{\max}, \\
&\frac{m_{\max}}{T_0}(1+d)^{1-m_{\max}}, &&m > m_{\max},
\end{aligned}
\r.
\end{equation}
where $m_{\max}$ is given in \eqref{Eq:mmax}. }
\end{proposition}
\vspace{-10pt}
For ease notation, the maximum computation rate,  $\mu({m_{\max}})$, is re-denoted as $\mu_{\max}$ hereafter. 

\vspace{-10pt}
\subsection{Computation Rates under Network Stability Constraint}\label{Subsection:StableProb}
\vspace{-10pt}
This subsection focuses on analzying the condition for the maximum computation rate of the typical CS to meet the network stability constraint in Definition \ref{Def:varepsilonStability}. The analysis combines the results from queueing theory, stochastic geometry and parallel  computing. The said constraint requires  $\rho$-fraction of mobiles, or equivalently $\rho$-fraction of CSs, to be stable, namely that  comp-latency is finite. According to queuing theory, stabilizing a typical  CS requires that the task-arrival rate $\Lambda$ should be strictly smaller than  the maximum departure rate $\mu^*(m_{\max})$: $\Lambda<\mu_{\max}$ \cite{BertsekasBook:DataNetwk:92}. Note that the former is a RV proportional to the random   number of mobiles, $N$,  connected to the typical CS while the latter is a constant. Then the stability probability $\mathsf{p}_s$ is given as
\begin{equation}
\mathsf{p}_s =\Pr[\Lambda<\mu_{\max}]=\Pr\l[N <\frac{\mu_{\max}}{\beta^*}\r],\label{Eq:StabilityConditionCS}
\end{equation}
where $\beta^*$ is the task-offloading rate given in \eqref{Eq:TaskOffloadingRate}.
It follows from  the network spatial model that  $N$ is a Poisson distributed RV with the mean $\bar{N}=(1-\delta)(1-\epsilon)\frac{\lambda_m}{\lambda_b}$. Using the distribution and \eqref{Eq:StabilityConditionCS} and applying Chernoff bound, we can obtain an upper bound on the maximum computation rate  required to meet the stability constraint as  shown below.
\vspace{-10pt}
\begin{proposition}[Computation Rates for  $\rho$-Stability]\label{lemma:chernoff} \emph{For the CSN to be  $\rho$-stable, a sufficient condition for the maximum computation rate of the typical CS is given as
\begin{align}\label{Eq:Chernoff}
\mu_{\max}\geq \bar{\Lambda}^*\cdot \exp\l(W\l(\frac{-\ln(\rho)}{\bar{N} e} - \frac{1}{e}\r)+1\r),
\end{align}
where  $W(\cdot)$ is the Lambert function,  the expected mobiles connected to the typical CS $\bar{N}=(1-\delta)(1-\epsilon)\frac{\lambda_m}{\lambda_b}$, and $\bar{\Lambda}^*$ represents the expected  arrival rate  given in \eqref{Eq:TaskArrivalRate}.
}
\end{proposition}
\vspace{-10pt}
\begin{proof}
See Appendix \ref{proof:chernoff}.
\end{proof}
\vspace{-5pt}

The above result shows that to satisfy the network-stability constraint, the maximum computation rate of each CS, $\mu_{\max}$,  should be larger than the expected task-arrival rate, $\bar{\Lambda}^*$,  scaled by a factor larger than one, namely   the exponential term  in \eqref{Eq:Chernoff}. Moreover, the factor grows as the stability probability $(1-\rho)$ increases. 

Last, it is useful for subsequent analysis to derive the expected arrival rate conditioned on that the typical CS is stable as shown below. 
\vspace{-10pt}
 \begin{lemma}[Expected Task-Arrival Rates for  Stable CSs]\label{Lemma:ConditionalExpectedArrival}\emph{Given that the typical CS is stable, the expected task-arrival rate is given as
\begin{align}\label{Eq:ConditionalExpectedArrival}
\E[\Lambda|\Lambda<\mu_{\max}]=\bar{\Lambda}^* \l(1-\frac{\Pr(N = \l\lfloor R\r\rfloor)}{1 - \rho}\r),
\end{align}
where  $R=\frac{\mu_{\max}}{\beta^*}$ measures  the maximum number of mobiles the CS can serve,  $\beta^*$ is the task-offloading rate per mobile in \eqref{Eq:TaskOffloadingRate}, $\bar{N}$ and $\bar{\Lambda}^*$ follow those in Proposition \ref{lemma:chernoff}, and the Poisson distribution function $\Pr(N = n )=\frac{\bar{N}^ne^{-\bar{N}}}{n!}$.}
\end{lemma}
\vspace{-10pt}
\begin{proof} See Appendix \ref{proof:lemma_taskArrival}.
\end{proof}


\vspace{-25pt}
\subsection{Expected Computation Latency}

In this subsection, the expected comp-latency, $\mathsf{T}_{\textrm{comp}}$,  is analyzed using the Markov chain in  Fig.~\ref{mmc_fig} and applying queueing theory. Exact analysis is intractable due to the fact that the departure rate $\mu(m)$ in the Markov chain is a non-linear function of state $m$. This difficulty is overcome by modifying   the Markov chain to give two versions corresponding to a M/M/m and a M/M/1 queues, yielding an upper and a lower bounds on $\mathsf{T}_{\textrm{comp}}$, respectively.

First, consider upper bounding $\mathsf{T}_{\textrm{comp}}$. To this end, the departure rate $\mu(m)$ in the Markov chain in Fig.~\ref{mmc_fig} with the following lower bound obtained by fixing all exponents as $(1-m_{\max})$:
\begin{align}\label{Eq:DepartureRateUpper}
\mu^-(m) = \begin{cases}
                \frac{m}{T_0}(1+d)^{1-m_{\max}}, ~~~1\leq m \leq m_{\max}, \\
                \frac{m_{\max}}{T_0}(1+d)^{1-m_{\max}}, ~~~m > m_{\max}.
             \end{cases}
\end{align}
As a result, the modified Markov chain is a M/M/$m_{\max}$ queue. The corresponding waiting time, denoted as $\mathsf{T}_{\textrm{comp}}^+$,
upper bounds $\mathsf{T}_{\textrm{comp}}$ since it reduces the computation rate. Applying classic results on M/M/m queues  (see e.g.,
\cite{BertsekasBook:DataNetwk:92}), the waiting time, $\mathsf{T}_{\textrm{comp}}^+$, for task arrival rate $\Lambda$ is 
\begin{align}
\mathsf{T}_{\textrm{comp}}^+(\Lambda) &=  \frac{m_{\max}}{\mu^-(m_{\max})} +  \frac{\tau\l(\frac{\Lambda}{\mu^-(m_{\max})}\r)^{m_{\max}}}{m_{\max}!\mu^-(m_{\max})\l( 1-\frac{\Lambda}{m_{\max}\mu^-(m_{\max})} \r)^2} , \label{Tcompa}
\end{align}
where  the coefficient $\tau$ is given as
\begin{align}
\tau = \l[\sum_{m=0}^{m_{\max}-1} \frac{1}{m!}\l(\frac{\Lambda}{\mu^-(m_{\max})}\r)^m + \sum_{m=m_{\max}}^{\infty}\frac{m_{\max}^{m_{\max}-m}}{m_{\max}!}\l(\frac{\Lambda}{\mu^-(m_{\max})}\r)^m\r]^{-1}. \label{ubrho}
\end{align}
Using
 \eqref{Eq:ConditionalExpectedArrival}, \eqref{Eq:DepartureRateUpper} and \eqref{Tcompa}, the upper bound  is given in the following theorem.
 \vspace{-10pt}
\begin{subtheorem}{theorem}
\begin{theorem}[Comp-Latency for Asynchronous Offloading]\label{ub_Tc_heavy_VMs}\emph{Consider asynchronous offloading. The average comp-latency  is upper bounded as
\begin{align}\label{tc_heavy_mmc_ub}
\mathsf{T}_{\textrm{comp}} \leq \frac{m_{\max}}{\mu_{\max}} + \l(\frac{m_{\max}}{\mu_{\max}}\r)^2\cdot \frac{\bar{\Lambda}^*}{(m_{\max}-1)!\l( m_{\max}-1 \r)^2}\cdot  \l(1-\frac{\Pr(N = \lfloor R\rfloor)}{1- \rho}\r),
\end{align}
where $ R$ follows that in Lemma \ref{Lemma:ConditionalExpectedArrival}, and $\bar{\Lambda}^*$ and $\mu_{\max}$ are specified in \eqref{Eq:TaskArrivalRate} and  \eqref{Eq:muopt}, respectively.
 }
\end{theorem}
\vspace{-10pt}
\begin{proof}
See Appendix \ref{proof:lemma_Tc_heavy_VMs}.
\end{proof}
Note that the positive factor $\l(1-\frac{\Pr(N =    \lfloor R\rfloor)}{1- \rho}\r)$ accounts for  Poisson distribution of mobiles.

Next, a  lower bound on $\mathsf{T}_{\textrm{comp}}$ is obtained as follows. One can observe from the Markov chain in Fig. \ref{mmc_fig} that for states $m \leq m_{\max}$, the departure rates are smaller than the maximum, $\mu_{\max}$. The reason is that for these states, there are not enough tasks for attaining  the maximum rate by  parallel computing. Then replacing all departure rates in the said Markov chain with the maximum $\mu_{\max}$
leads to a lower bound on $\mathsf{T}_{\textrm{comp}}$. The resultant Markov chain  corresponds to a M/M/1 queue. Then using the modified Markov chain and  the well-known results from M/M/1 queue (see e.g.,
\cite{BertsekasBook:DataNetwk:92}), the comp-latency for   given arrival rate $\Lambda$ can be lower bounded as
\begin{align}
\mathsf{T}_{\textrm{comp}}(\Lambda) \geq\frac{1}{\mu_{\max}-\Lambda}.
\end{align}
By taking expectation over $\Lambda$ and applying Jensen's inequality,
\begin{eqnarray}\label{Eq:ExpectedTcomp}
\mathsf{T}_{\textrm{comp}}= \E[\mathsf{T}_{\textrm{comp}}(\Lambda)]
\geq \E\l[\l.\frac{1}{\mu_{\max}-\Lambda}\r\rvert \Lambda<\mu_{\max}\r]
\geq \frac{1}{\mu_{\max}-\E[\Lambda|\Lambda<\mu_{\max}]}.
\end{eqnarray}
Using \eqref{Eq:ExpectedTcomp} and  Lemma \ref{Lemma:ConditionalExpectedArrival}, we obtain the following result.

\vspace{-10pt}
\begin{theorem}[Comp-Latency for Asynchronous Offloading]\label{lb_Tc_heavy_VMs}\emph{Consider asynchronous offloading. The average comp-latency  is lower bounded as
\begin{align}\label{tc_heavy_mmc_lb}
\mathsf{T}_{\textrm{comp}}\geq  \frac{1}{\mu_{\max} - \bar{\Lambda}^*\cdot\l(1-\frac{\Pr(N =    \lfloor R\rfloor)}{1- \rho}\r)},
\end{align}
where $ R$ follows that in Lemma \ref{Lemma:ConditionalExpectedArrival}, and $\bar{\Lambda}^*$ and $\mu_{\max}$ are specified in \eqref{Eq:TaskArrivalRate} and  \eqref{Eq:muopt}, respectively.
}
\end{theorem}
\end{subtheorem}
\vspace{-20pt}

\begin{remark}[Computation-Resource  Provisioning]\emph{Consider a MEC network provisioned with sufficient computation resources, 
$\mu_{\max}/\bar{\Lambda}^*\gg 1$. It follows from  Theorem \ref{lb_Tc_heavy_VMs}
\begin{align}\nn
\mathsf{T}_{\textrm{comp}}\geq  \frac{1}{\mu_{\max}}\left(1 + \frac{c_1\bar{\Lambda}^*}{\mu_{\max}}\r),
\end{align}
where $c_1$ is a constant. This lower bound   has  a similar form as  the upper bound in Theorem \ref{ub_Tc_heavy_VMs}. From these results, one can infer that the comp-latency for asynchronous offloading can be approximated written in the following form: 
\begin{align}\label{Eq:Tcomp:Re}
\boxed{
\mathsf{T}_{\textrm{comp}}\approx  \frac{c_2}{\mu_{\max}}\left(1 + \frac{c_3\bar{\Lambda}^*}{\mu_{\max}}\r), \qquad \frac{\mu_{\max}}{\bar{\Lambda}^*}\gg 1,
}
\end{align}
where $\{c_2, c_3\}$ are constants. The result suggests that to contain comp-latency, the provisioning of computation resources for the MEC network must consider two factors. First of all, the maximum computation rate, $\mu_{\max}$, for each CS must be sufficient large. At the same time, the computation rate  must scale linearly with the total arrival rate such that the computation resource allocated for a single  offloaded task, measured by the ratio  $\mu_{\max}/\bar{\Lambda}^*$, is sufficiently large. 
 }
\end{remark}
\vspace{-30pt}
\subsection{Energy Efficiency}
\vspace{-10pt}
Based on the above analytical results so far, the subsection tempts to discuss the energy savings of offloading than local computing. First, the energy consumption of  offloading, denoted by $\mathsf{E}_{\mathrm{off}}$, can be derived via multiplying mobile's transmission power $P$ by the offloading duration $G^* \mathsf{T}_{\min}$. 
To satisfy the minimum average signal strength at the boundary of the MEC service zone, 
$P$ should scale with the  radius ${r_0}$ as 
$P\propto r_0^{\alpha}$. Recalling $r_0\propto\lambda_b^{-\frac{1}{2}}$ and $\mathsf{T}_{\min}\propto\ell $ where $\ell$ is the task size, the resultant energy consumption of $\mathsf{E}_{\mathrm{off}}$ is given as
\begin{align}\label{Energy_off}
\mathsf{E}_{\mathrm{off}}=c_4 \frac{G^*  \ell}{{\lambda_b}^{\frac{\alpha}{2}}},
\end{align}
where $c_4$ is a constant depending on the minimum signal strength and $\mathsf{T}_{\min}$. Next, it is well studied in\cite{Zhang2013TVT} that the optimal $\mathsf{E}_{\textrm{loc}}$ is proportional to ${\ell}^3$, and inversely proportional to the square of the deadline requirement which could be set as the total latency $\mathsf{T}_{\textrm{comm}}+\mathsf{T}_{\textrm{comp}}$ without loss of generality.  
Thus, $\mathsf{E}_{\textrm{loc}}$ is given as
\begin{align}\label{Energy_Local}
\mathsf{E}_{\textrm{loc}}=c_5 \frac{{\ell}^3}{(\mathsf{T}_{\textrm{comm}}+\mathsf{T}_{\textrm{comp}})^2},
\end{align}
where $c_5$ is a constant depending on the chip architecture. where $c_5$ is a constant depending on the chip architecture. 
As a result, the condition of energy savings, namely $\mathsf{E}_{\mathrm{off}}<\mathsf{E}_{\textrm{loc}}$, is given in terms of $\ell$ and $\lambda_b$ as
\begin{align}\label{EnergySavingsCondition}
\ell^2 \lambda_b^{\frac{\alpha}{2}}>\frac{c_4}{c_5}{G^*}\l(\mathsf{T}_{\textrm{comm}}+\mathsf{T}_{\textrm{comp}}\r)^2.
\end{align}
It is observed that the right-side of \eqref{EnergySavingsCondition} is dominantly affected by the expected number of mobiles $\frac{\lambda_m}{\lambda_b}$ (see \eqref{Eq:Tcomm}, \eqref{Eq:OptimalGApprox}, \eqref{Eq:QuasiConcaveLambda} and \eqref{Eq:Tcomp:Re}). In other words, given a mobile density $\lambda_m$ and the task size $\ell$, there exists the minimum density of APs $\lambda_b$ to satisfy the condition in \eqref{EnergySavingsCondition}. Then under the condition in \eqref{EnergySavingsCondition}, the energy savings due to offloading is given as $\mathsf{E}_{\mathrm{loc}}-\mathsf{E}_{\mathrm{off}}$ with $\mathsf{E}_{\mathrm{off}}$ and $\mathsf{E}_{\mathrm{loc}}$ given in \eqref{Energy_off} and \eqref{Energy_Local}, respectively.

\vspace{-20pt}
\subsection{MEC Network Provisioning and Planning}
\vspace{-5pt}
 Combining the results from the preceding analysis on comm-latency and comp-latency yields some guidelines for the provisioning and planning of a MEC network as discussed below. Assume that the network is required to support computing for mobiles with density $\lambda_m$ with targeted  expected comm-latency $\mathsf{T}_{\textrm{comm}}$ and comp-latency  $\mathsf{T}_{\textrm{comp}}$. The network resources are quantified by the bandwidth $B$, the density of AP (or CS) $\lambda_b$, the maximum computing rate of each CS $\mu_{\max}$. 

First, consider the planning of the RAN. Combining the above results suggest the following guidelines for network provisioning and planning. As shown in Section \ref{Subsection:TaskArrivalRate}, under the network-coverage constraint $(1-\epsilon)$, the expected task-arrival rate at a AP, representing the RAN offloading throughput, is a quasi-concave function of the expected number of mobiles per AP, $\lambda_m/\lambda_b$, with a global maximum. Therefore, given a mobile density $\lambda_m$, the AP density should be chosen for maximizing the RAN offloading throughput. Next, based on results in Theorem~\ref{theo:optimalG} and \eqref{Eq:TaskOffloadingRate}, sufficient large channel  bandwidth $B$  should be provisioned to achieve the targeted   $\mathsf{T}_{\textrm{comm}}$ for given mobile and AP densities, mobile task-generation rates, and task sizes. 

Next, consider the planning of the CSN. Under the  network-stability constraint, the maximum CS computing rate for parallel computing  should be  planned to be larger than the expected task-arrival rate scaled by a factor larger than one, which is determined by the allowed fraction of unstable CSs (see Proposition \ref{lemma:chernoff}). Then,  the maximum computing rate should be further planned to achieve the targeted  $\mathsf{T}_{\textrm{comp}}$ for computing an offloaded task using Theorems \ref{ub_Tc_heavy_VMs} and~\ref{lb_Tc_heavy_VMs}.  
\vspace{-15pt}
\section{Computation Latency Analysis: Synchronous Offloading}\label{Section:CompLat_syn}
\vspace{-5pt}

In the preceding section, the process of asynchronous task arrival at a CS can be approximated using a Markov chain, allowing tractable analysis of comp-latency using theories of M/M/$m$ and M/M/$1$ queues. This approach is inapplicable for synchronous offloading and the resultant periodic task arrivals at the CS. Though tractable  analysis in general is difficult, it is possible for two special cases defined as follows.
\vspace{-10pt}
\begin{definition}[Special Cases: Light-Traffic  and Heavy-Traffic]\label{def:lightHeacy}\emph{
A \emph{light-traffic} case refers to one that the task-arrival rate  is much smaller than the computation rate such that  the  queue at the CS  is always empty as observed by a new arriving task. In contrast, a \emph{heavy-traffic} case  refers to one that the task-arrival rate  is close to  the computation rate such that there are always at least  $m_{\max}$ tasks in the queue.}
\end{definition}
\vspace{-10pt}
The comp-latency for these two special cases are analyzed  to give insights into the performance of CSN with underloaded CSs and those with overloaded CSs.

\vspace{-20pt}
\subsection{Expected Computation Latency with Light-Traffic }
\vspace{-5pt}
First, the dynamics of the task queue at the typical CS is modelled as follows.  Recall that task arrivals are periodical, occurring at the beginning of every frame. Consider the typical CS. Let $\mathcal{Q}_t$ and $\mathcal{A}_t$ denote  be the numbers of existing and arriving tasks at the beginning of frame~$t$, respectively, and $\mathcal{C}_t$ the number of departing tasks during frame $t$. Then the evolution  of $\mathcal{Q}_t$  can be described mathematically as
\begin{equation}\label{Eq:DynamicEquation}
\mathcal{Q}_{t+1} = \max\l[ \mathcal{Q}_t + \mathcal{A}_{t+1} - \mathcal{C}_t, 0 \r].
\end{equation}
The general  analysis of comp-latency using \eqref{Eq:DynamicEquation} is difficult. The main difficulty lies in deriving the distribution of $\mathcal{C}_t$ that depends on the number of VMs that varies continuously in time since the computation time for simultaneous tasks are random and inter-dependent. To overcome the difficulty, consider the case of light-traffic where  a number of offloaded tasks arrives at the typical CS to see an empty queue and an idling server. Correspondingly, the evolution equation in \eqref{Eq:DynamicEquation} is modified such that given $\mathcal{A}_{t+1}\neq 0 $, $\mathcal{Q}_t = \mathcal{C}_t = 0 $, yielding   $\mathcal{Q}_{t+1} =  \mathcal{A}_{t+1}$.

Next, given this simple equality, deriving the expected comp-latency reduces to analyzing the latency for computing a random number of $\mathcal{A}$ tasks at the CS, which arrives at the beginning of an arbitrary frame. Without loss of generality, the tasks are arranged in an ascending order in terms of computation time and referred to as Task $1, 2, \cdots, \mathcal{A}$. Moreover, let  $\mathcal{L}_n$ denote the expected  computing time for Task $n$ and hence $\mathcal{L}_1 \leq \mathcal{L}_2 \leq \cdots \leq \mathcal{L}_{\mathcal{A}}$. Then the expected  comp-latency $\mathsf{T}_{\textrm{comp}}$ can be  written in terms of $\{\mathcal{L}_n\}$ as
\begin{align}
\mathsf{T}_{\textrm{comp}} = \E_{\mathcal{A}}\l[\l.\frac{\sum_{n=1}^{{\mathcal{A}}} {\mathcal{L}_n}  }{\mathcal{A}}\r\rvert \mathcal{A}>0 \r]. \label{syn_tc_light}
\end{align}

To obtain bounds on $\mathsf{T}_{\textrm{comp}}$ in closed form, a useful result is derived as follows.  Given $m$ VMs, recall that the computation time of a task follows the exponential distribution with the mean being the inverse of the computation rate  $\mu(m)$ in \eqref{Eq:muopt}.  Using the memoryless property of exponential distribution, a useful relation between $\{\mathcal{L}_n\}$ is obtained as
\begin{align}\label{Eq:RecursiveL}
\mathcal{L}_n =\mathcal{L}_{n-1}+ \frac{1}{\mu(m)}
=
\l\{
\begin{aligned}
&\mathcal{L}_{n-1}+ \frac{1}{\mu_{\max}}, && 1\leq n\leq \mathcal{A}-m_{\max}+1,\\
&\mathcal{L}_{n-1} +\frac{1}{\mu(\mathcal{A} - n +1)}, && \text{otherwise},
\end{aligned}
\r.
\end{align}
with  $\mathcal{L}_0=0$. Note that $\mu(1) \leq \mu(m) \leq \mu_{\max}$ for all $m$. Thus, it follows from \eqref{Eq:RecursiveL} that
\begin{equation}
\frac{n}{\mu_{\max}} \leq \mathcal{L}_n \leq \frac{n}{\mu(1)}. \label{Eq:LInequality2}
\end{equation}
Substituting \eqref{Eq:LInequality2} into \eqref{syn_tc_light} gives
\begin{align}
\frac{1}{\mu_{\max}}\cdot \E_{\mathcal{A}}\l[\l.\frac{\sum_{n=1}^{{\mathcal{A}}} n  }{\mathcal{A}}\r\rvert \mathcal{A}>0 \r]\leq \mathsf{T}_{\textrm{comp}} \leq \frac{1}{\mu(1)} \cdot \E_{\mathcal{A}}\l[\l.\frac{\sum_{n=1}^{{\mathcal{A}}} n  }{\mathcal{A}}\r\rvert \mathcal{A}>0 \r].
\end{align}
Recalling the number of arriving tasks $\mathcal{A}$ follows a Poisson RV with mean $\bar{A}^*$ of \eqref{Eq:AverageTaskArrival},  
bounds on the comp-latency is  obtained shown in the following theorem.
\vspace{-10pt}
\begin{theorem}[Comp-Latency for Synchronous Offloading]\label{lb_Tc_light_VMs_syn}\emph{Consider the case of synchronous offloading with \emph{light-traffic}. The expected  comp-latency can be  bounded as
\begin{equation}\nn
\frac{1}{2\mu_{\max}}\l(1+\frac{\bar{A}^*}{1-e^{-\bar{A}^*}}\r)\leq \mathsf{{T}}_{\textrm{comp}} \leq \frac{1}{2\mu(1)}\l(1+\frac{\bar{A}^*}{1-e^{-\bar{A}^*}}\r).
\end{equation}
where
$\bar{A}^*$ is the expected number of arriving tasks to the typical CS per frame given in \eqref{Eq:AverageTaskArrival}.
}
\end{theorem}
\vspace{-20pt}
\begin{remark}[Comparison with Asynchronous Offloading]\emph{
From the results in the above theorem, one can infer that for the current case, the comm-latency can be approximated as 
\begin{equation}
\boxed{
\mathsf{{T}}_{\textrm{comp}} \approx  \l\{
\begin{aligned}
&\frac{1}{\mu(1)}, && \bar{A}^* \rightarrow 0,\\
&\frac{\bar{A}^*}{2\mu(m_{\max})},&& \bar{A}^* \gg 1.
\end{aligned}
\r.
}
\label{Eq:UpperBoundLightTraffic}
\end{equation}
Comparing the expression with the counterpart  for asynchronous offloading in \eqref{Eq:Tcomp:Re}, it is unclear which case leads to longer comp-latency. However,  simulation shows that in general, synchronizing offloading tends to incur longer latency by overloading CSs and thereby suffering more from I/O interference in parallel-computing.}
\end{remark} 

\vspace{-25pt}
\subsection{Expected Computation Latency with Heavy-Traffic} \label{sec:heacy}

This subsection focuses on analyzing the expected comp-latency, $\mathsf{T}_{\textrm{comp}}$,  for  the case of heavy-traffic as defined in Definition \ref{def:lightHeacy}. For this case, with the queue being always non-empty, the equation in \eqref{Eq:DynamicEquation}  describing the queue evolution reduces to $\mathcal{Q}_{t+1} = \mathcal{Q}_t + \mathcal{A}_t - \mathcal{C}_t$. The key step in deriving $\mathsf{T}_{\textrm{comp}}$ is to apply the said equation to the analysis of the expected queue length. The technique involves taking expectation of the squares of the two sides of the equation as follows:
\begin{align}\label{asy_heacy_dyeq_square}
 \E\l[ \mathcal{Q}^2_{t+1} \r]= \E\l[ \(\mathcal{Q}_t + \mathcal{A}_t - \mathcal{C}_t\r)^2\r]
&=\E\l[ \mathcal{Q}^2_t \r]+\E\l[\l( \mathcal{A}_t - \mathcal{C}_t \r)^2\r]+2\E\l[{\mathcal{Q}}_t (\mathcal{A}_t - \mathcal{C}_t) \r].
\end{align}
Since $\mathcal{Q}_t$, $\mathcal{A}_t$ and $\mathcal{C}_t$ are independent of  each other and $\E\l[ \mathcal{Q}^2_{t+1} \r]=\E\l[ \mathcal{Q}^2_t\r]$ given the stable CS, 
\begin{align}
\E\l[ \mathcal{Q}\r] = \frac{\E\l[\mathcal{A}^2\r] + \E\l[\mathcal{C}^2\r] - 2\E\l[\mathcal{A}\r]\E\l[\mathcal{C}\r]}{2\l(  \E\l[\mathcal{C}\r] - \E\l[\mathcal{A}\r] \r)}, \label{EQ_Heavy}
\end{align}
where the subscripts $t$ of  $\mathcal{Q}_t$, $\mathcal{A}_t$ and $\mathcal{C}_t$ are omitted to simplify notation.
Given the number of connected mobiles, $N$, the number of arrival tasks $\mathcal{A}$ follows a Poisson distribution
with the first and second moments being  $\E\l(\mathcal{A}\mid N \r)={N} p_L^*$ and $\E\l(\mathcal{A}^2\mid N \r)=N p_L^*+(Np_L^*)^2$
respectively, where the task-offloading probability $p_L^*$ is given in \eqref{Eq:TaskOffloadingProb}. Next, under the heavy-traffic assumption, the total computation rate of the CS is $\mu_{\max}$. It follows that the departure process at the typical CS is Poisson distributed where  the first and second moments are  $\E[\mathcal{C}]= \mu_{\max}L^*$ and $\E[\mathcal{C}^2]= \mu_{\max}L^*+[ \mu_{\max}L^*]^2$,  respectively. Substituting the results into \eqref{EQ_Heavy} gives
\begin{align}
\E\l[ \mathcal{Q}\mid N \r] =\frac{{N} p_L^*}{\mu_{\max}L^*-{N} p_L^*}+\frac{1}{2}\cdot \l(\mu_{\max}L^*-{N} p_L^*\r)+\frac{1}{2}. \label{EQ_Heavy:a}
\end{align}
In addition, to satisfy the condition for stabilizing the  CS, the arrival rate $\E\l[\mathcal{A}\r]$ should be strictly  smaller than  the departure rate $\E[\mathcal{C}]$. This places  a constraint on the maximum of $N$, namely that $N \leq \l\lfloor R \r\rfloor$ with $ R$ defined in Lemma \ref{Lemma:ConditionalExpectedArrival}. Under this constraint,  applying Little's theorem obtains the expected  comp-latency  $\mathsf{T}_{\textrm{comp}}$ as
\begin{align}
\mathsf{T}_{\textrm{comp}}& =  \l\{\E\l[\l. \frac{\E\l[ \mathcal{Q}\mid N \r]}{\E\l[ \mathcal{A}\mid N \r]}\r\rvert
N \leq \l\lfloor R \r\rfloor
\r]-\frac{1}{2}\r\}\cdot L^*. \label{tc_heacy_syn} 
\end{align}
Combining \eqref{EQ_Heavy:a} and \eqref{tc_heacy_syn} yields the main result of this sub-section as shown below.
\vspace{-10pt}
\begin{theorem}[Comp-Latency for Synchronous Offloading]\label{lb_Tc_heavy_VMs_syn}\emph{Consider the case of synchronous offloading with  heavy-traffic. The expected  comp-latency  is given as
\begin{align}\label{ratio_heavy_mmc_ub}
\mathsf{T}_{\textrm{comp}}=\l\{\frac{1}{p_L^*}\E\l[\frac{1}{ R-N}\r] +\frac{1}{2}\l( R + \frac{1}{p_L^*}\r) \E\l[\frac{1}{N}\r] - 1\r\}L^*,
\end{align}
where the constant $ R$ and the distribution of $N$ follow those in Lemma \ref{Lemma:ConditionalExpectedArrival}. 
}
\end{theorem}
\vspace{-20pt}

\begin{remark}[Comparison with Asynchronous Offloading]\label{remark:ComparingAsyncHeavy}\emph{
By applying  Jensen's inequality, the comp-latency of \eqref{ratio_heavy_mmc_ub} can be lower bounded as 
\begin{align}
\mathsf{T}_{\textrm{comp}}&\geq \frac{1}{\mu_{\max}-c_4\bar{\Lambda}^*}+\frac{L^*}{2}\cdot \frac{\mu(m_{\max})}{c_4\bar{\Lambda}^*}+\frac{1}{c_4\bar{\Lambda}^*}- L^*,
\end{align}
where  $c_4$ is a constant. Since for the case of heavy-traffic, the task-arrival rate $c_4\bar{\Lambda}^*$ approaches the maximum computation rate $\mu_{\max}$, 
\begin{align}
\boxed{
\mathsf{T}_{\textrm{comp}}\geq \frac{1}{\mu_{\max}-c_4\bar{\Lambda}^*},\qquad c_4\bar{\Lambda}^*\rightarrow \mu_{\max}. 
}
\end{align}
The above lower bound has the same form as the asynchronous-offloading counterpart in \eqref{tc_heavy_mmc_lb}. Both diverge as the task-arrival rate approaches the maximum computation rate. 
}
\end{remark}
\vspace{-20pt}

\section{Simulation Results}\label{sim}
\vspace{-10pt}
In this section, analytical results on  comm-latency and comp-latency are evaluated by simulation. The simulation parameters have the following default settings  unless specified otherwise. The densities of APs and mobiles are $\lambda_b = 2 \times 10^{-2} \ \mathrm{m}^{-2}$ and $\lambda_m = 5 \times 10^{-2} \ \mathrm{m}^{-2}$, respectively. The $\SIR$ threshold is set as $\theta = 1$ dB and the path-loss exponent is $\alpha = 3$. For the network coverage parameter $\delta$ is  $\delta = 10^{-2}$, corresponding to the radius of MEC service zone being  $r_0 = 12 \mathrm{m}$. The total bandwidth is $B = 6$ MHz. The data size per task is fixed as $\ell = 0.5 \times 10^{6}$ bits. The single-task computation time $T_0$ in the parallel-computation model is set as  $T_0 = 0.1$ (sec) and the factor arising from I/O interference is  $d = 0.2$. The task generation probability per slot  is $p = 0.2$. The parameters  $\epsilon$ and $\rho$ are both set as $0.05$.

\begin{figure}[t]
\centering
\subfigure[Effect of mobile density.]{\includegraphics[width=8cm]{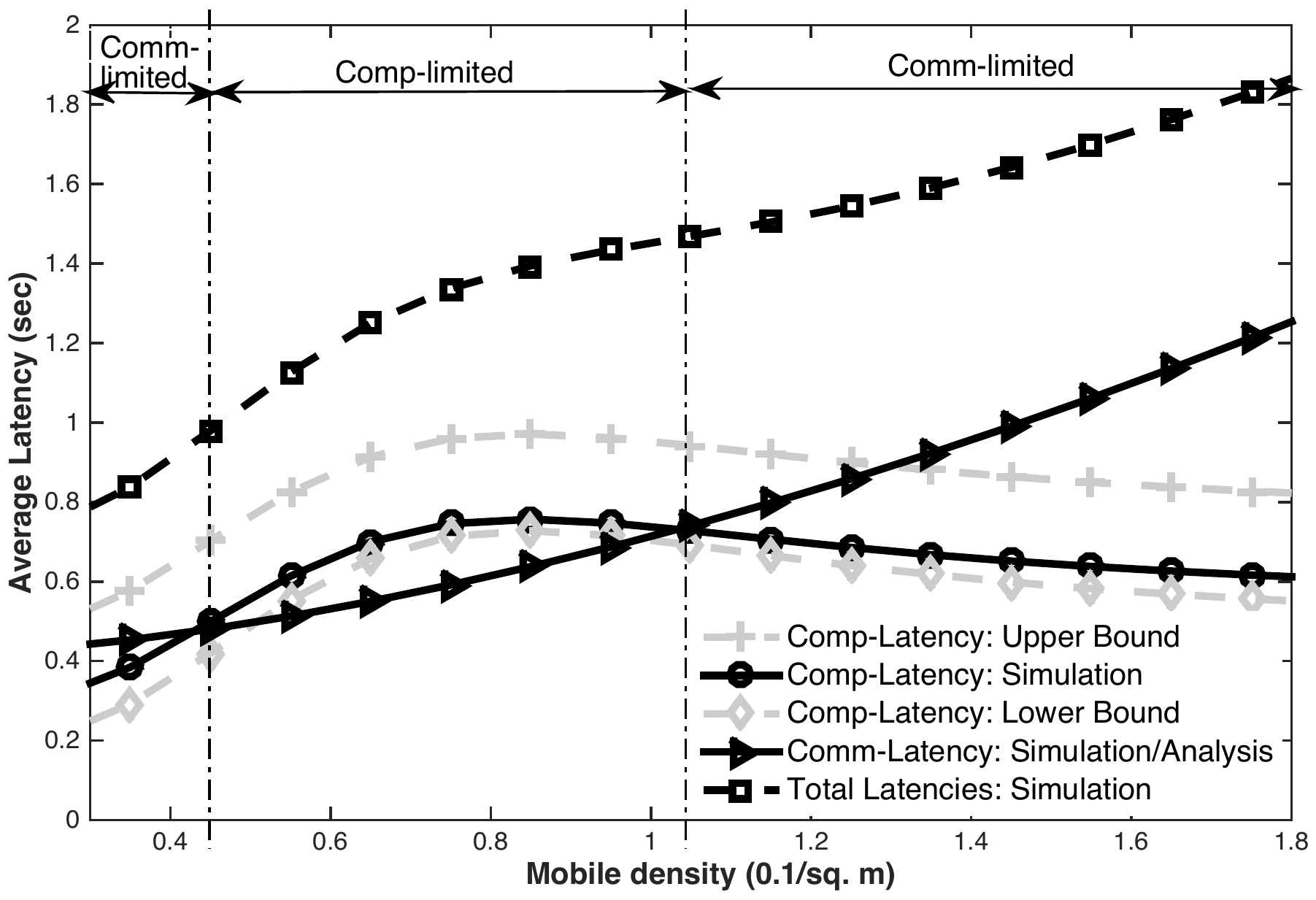}}
\subfigure[Effect of task generating rate.]{\includegraphics[width=8cm]{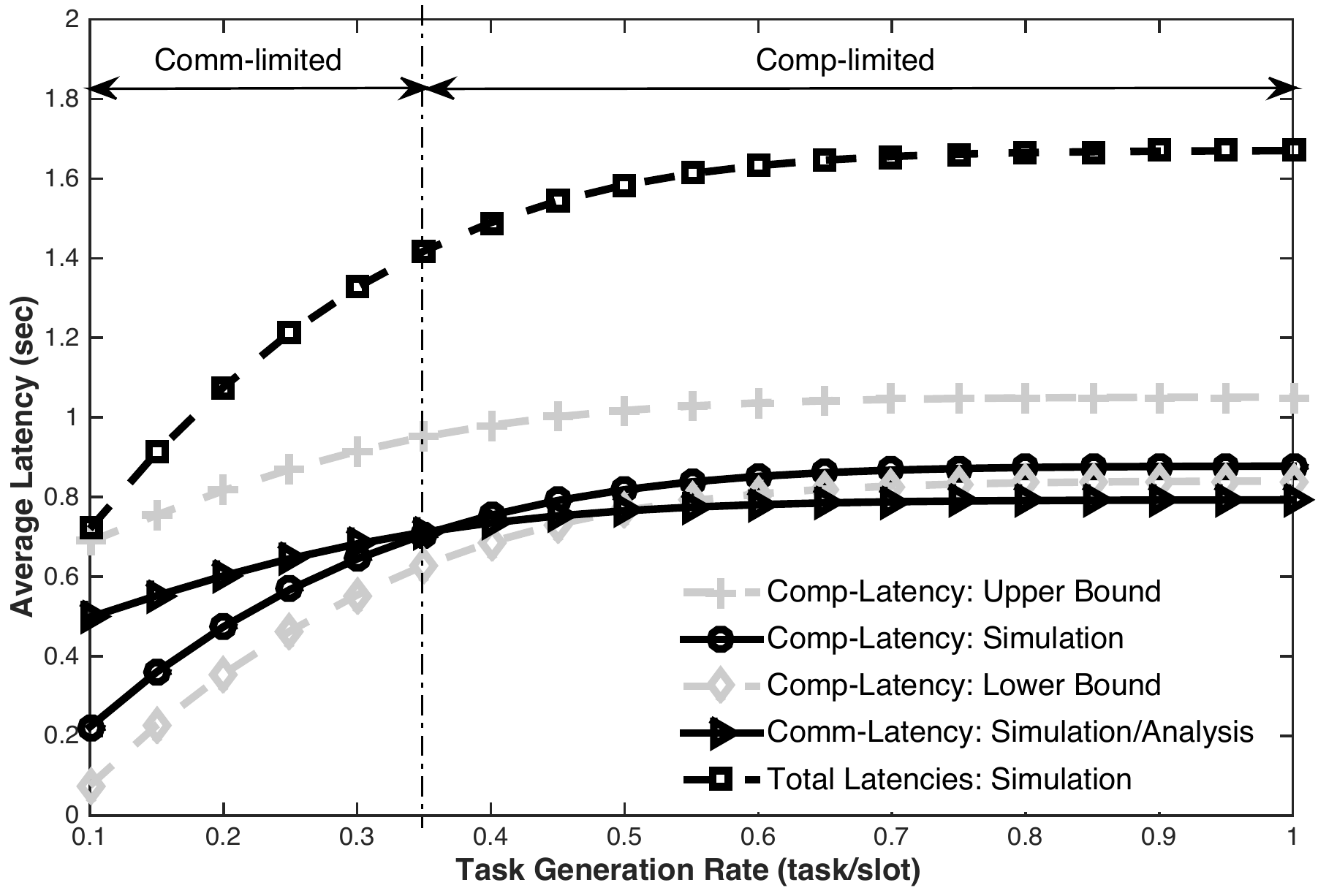}}
\vspace{-20pt}
\caption{Comparisons between comm-latency and comp-latency for the case of  asynchronous offloading.}
\label{fig_asyLaten}
\vspace{-30pt}
\end{figure}

We present the comparisons between Monte Carlo simulations ($10^4$ realizations) and analytical results in all figures. For each realization, both mobiles and APs are distributed in the plane based on the PPPs.  Each mobile randomly generates an offloading task and transmits it to its corresponding AP. Then the AP generates VMs and performs the computation upon the task arrivals. A queue of tasks will appear if the task arrival rate is large than the computation rate (e.g., too many tasks arrived at the AP at the same time). The VM will be released when the corresponding task is computed.

Fig.~\ref{fig_asyLaten} compares expected  comm-latency and comp-latency for the case of asynchronous offloading. The effects of  mobile density $\lambda_m$ and task generating rate $p$ are investigated and several observations can be made. As shown in Fig.~\ref{fig_asyLaten}(a), the expected comp-latency as a function of the mobile density  is observed to exhibit the \emph{quasi-concavity} described in Remark~\ref{remark:depOptTcomMobile}. In contrast, the expected comm-latency is a monotone increasing function following the scaling law in Remark~\ref{remark:commScaLaw}. These properties lead to the partitioning of the range of mobile density into three network-operation regimes as indicated in Fig.~\ref{fig_asyLaten}(a). In particular, the middle range corresponds to comp-limited regime while others are comm-limited. 
Next, consider the effect of task-generation rate at mobiles, specified by the task-generation probability $p$.  Both types of latency are observed to converge to corresponding limits as the rate  grows. Their different scaling laws result in the partitioning of the range of task-generation rate into comm-limited and comp-limited regimes. Last, one can observe from both figures that the lower bound on comp-latency as derived in \eqref{tc_heavy_mmc_lb} is tighter than the upper bond therein.  

Fig.~\ref{fig_synLaten} compares expected  comm-latency and comp-latency for the case of synchronous offloading. The same observations in the case of synchronous offloading also apply in the current case  except that the quasi-concavity of the expected comp-latency with respect to mobile density is not shown in the considered range.  Some new observations can be made as follows. Comparing Fig.~\ref{fig_asyLaten} and \ref{fig_synLaten} shows that synchronizing offloading results in longer comp-latency. Next, the center of the comp-limited range in Fig.~\ref{fig_synLaten}(a) corresponds to the case of heavy-traffic studied in Section~\ref{sec:heacy}. Consequently, the derived upper bound on expected comp-latency for this case is tight. For other ranges of mobile density, the bounds derived for the case of light traffic are tighter. Last, Fig.~\ref{fig_synLaten}(b) shows that the expected comp-latency is  tightly approximated  by bounds derived for the light-traffic case when the task-arrival rate is  small ($\leq 0.3$) and by that for the heavy-traffic case when the rate is  large ($> 0.7$), validating the results.


\begin{figure}[t]
\centering
\subfigure[Effect of mobile density.]{\includegraphics[width=8cm]{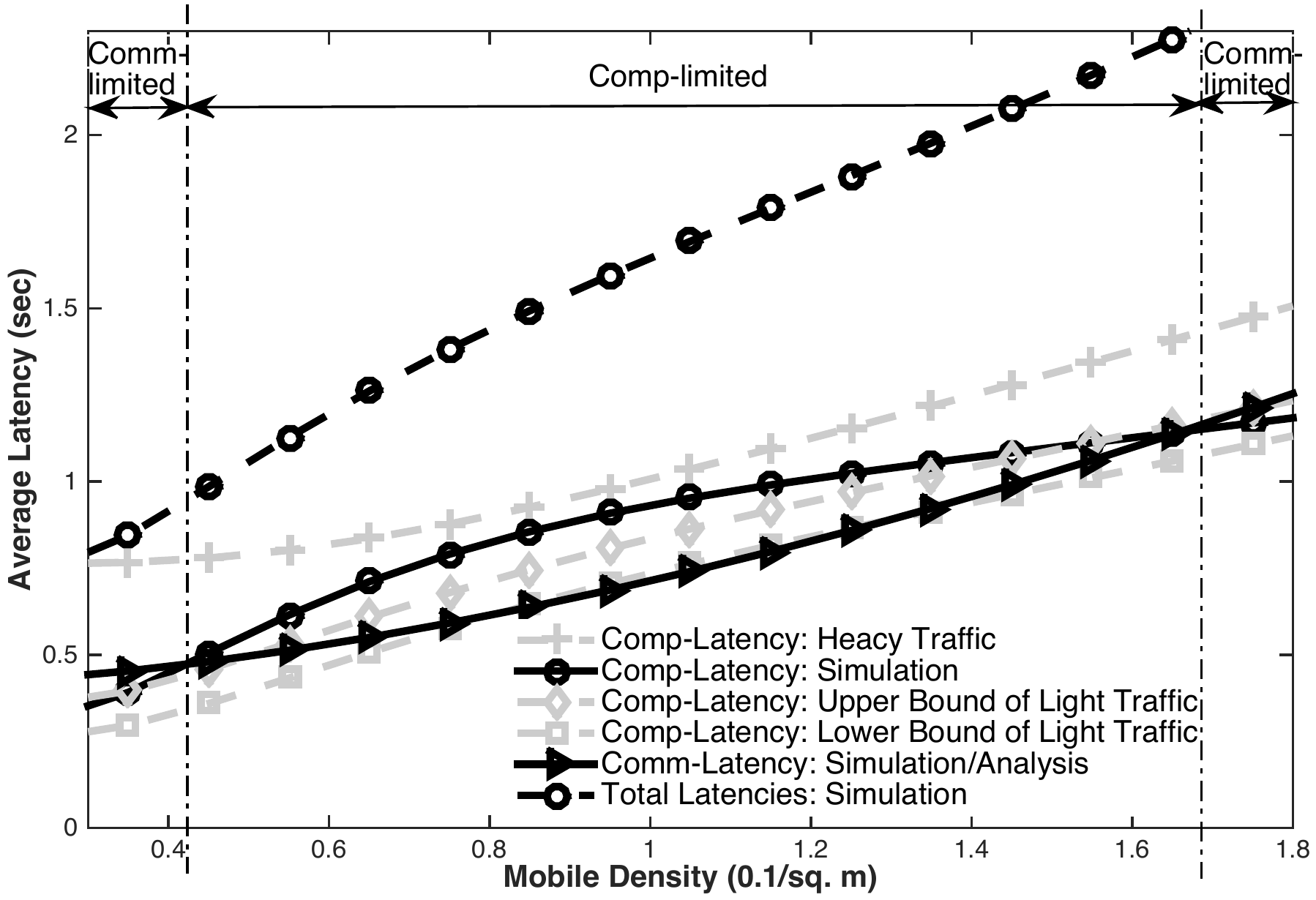}}
\subfigure[Effect of task generating rate.]{\includegraphics[width=8cm]{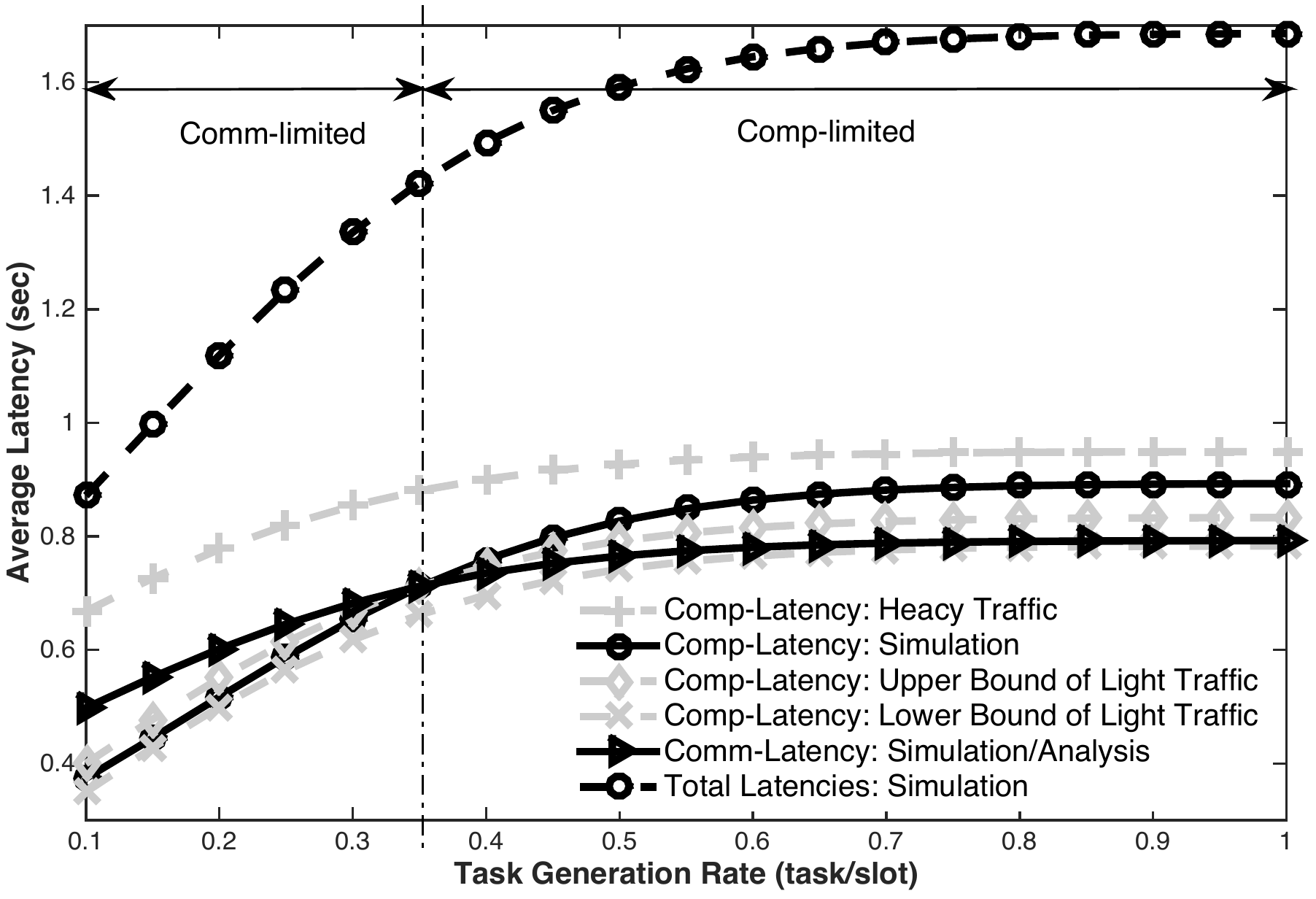}}
\vspace{-20pt}
\caption{Comparisons between comm-latency and comp-latency   for the case of  synchronous offloading.}
\label{fig_synLaten}
\vspace{-30pt}
\end{figure}

%

\vspace{-20pt}
\section{Conclusion Remarks}\label{con}
\vspace{-10pt}

{In this work, we have first studied the network-constrained latency performance of a large-scale MEC network, namely comm-latency and comp-latency 
under the constraints of RAN connectivity and CSN stability.
To study the tradeoffs between these metrics and constraints and model the cascaded architecture of RAN and CSN, 
the MEC network has been modelled using stochastic geometry featuring diversified aspects of wireless access and computing.
Based on the model, the average comm-latency and comp-latency have been analyzed by applying the theories of stochastic geometry, queuing and parallel computing. 
In particular, their scaling laws have been derived with respect to various network parameters ranging from the densities of mobiles and APs to the computation capabilities of CSs.
The results provide useful guidelines for MEC-network provisioning and planning to avoid either the RAN or CSN being a performance  bottleneck.}

{The current work can be extended in several directions. 
In this work, we consider a single type of computation task of which the task size and the average computation time are identical. However, considering different types of tasks makes the latency analysis more challenging but is of practical relevance. 
Next,  studying a large-scale hierarchical fog computing network comprising mobiles, edge cloud and central cloud is aligned with recent advancements in edge computing.
Last, considering advanced techniques such as VM migration and cooperative computing to reduce the latency will be another promising direction.}

\vspace{-20pt}
\appendix

\vspace{-10pt}

\subsection{Proof of Lemma \ref{lemma:exitsUniG}}\label{proof:xiG}
\vspace{-10pt}
The first derivative of $\xi(G)$ for $G$ given in \eqref{xi} can be derived as:
\begin{align}
\frac{\partial \xi(G)}{\partial G} = -\frac{\Delta}{\alpha}\l( G^{-\frac{2 + \alpha}{\alpha}}\l( \alpha G \mathsf{T}_{\min} \ln(1-p) (1-p)^{G \mathsf{T}_{\min}} - 2(1-p)^{G \mathsf{T}_{\min}} + 2 \r) \r), \label{firtsDerXi}
\end{align}
where $\Delta = \frac{2(1-\delta)\ln \delta^{-1}}{\alpha} \mathcal{B}( \alpha ) \l( \frac{\lambda_m}{\lambda_b}\r) \theta^{\frac{2}{\alpha}}$. The existence of $g_0$ is easily proved because \eqref{firtsDerXi} is strictly positive and negative when $G \rightarrow 0$ and $G \rightarrow \infty$ respectively.  Next, the solution of $g_0$ to solve  $\frac{\partial \xi(G)}{\partial G}   = 0$ is $g_0 =\frac{\alpha W\l( -\frac{2}{\alpha}e^{-\frac{2}{\alpha}} \r) + 2}{\alpha \mathsf{T}_{\min} \ln(1-p)}$,
where $W(x)$ is the Lambert function. Since the value inside the Lambert function is negative, there are two candidates for $g_0$: one is from the principle branch of Lambert function $W\l( -\frac{2}{\alpha}e^{-\frac{2}{\alpha}} \r)$, and the other is the lower branch $W_{-1}\l( -\frac{2}{\alpha}e^{-\frac{2}{\alpha}} \r)$. The principle branch makes $g_0=0$, but the lower branch satisfies $g_0>0$, completing the proof.

\vspace{-20pt}
\subsection{Proof of Proposition \ref{lemma:chernoff}}\label{proof:chernoff}
\vspace{-10pt}
Applying Chernoff bound on $\mathsf{p}_s$ in \eqref{Eq:StabilityConditionCS} makes
\begin{align}
\mathsf{p}_s 
&\geq 1- \exp\l(- \frac{\mu_{\max}}{\beta^*} \ln\l(\frac{ \mu_{\max} }{\bar{N} \beta^*}\r)  +\frac{\mu_{\max}}{\beta^*}-\bar{N}\r)
\geq 1 - \rho,
\end{align}
which is equivalent to $\mu_{\max}\geq \bar{\Lambda}^*\cdot \exp\l(W\l(\frac{-\ln(\rho)}{\bar{N} e} - \frac{1}{e}\r)+1\r)$ as Proposition \ref{lemma:chernoff} shows.

\vspace{-20pt}
\subsection{Proof of Lemma \ref{Lemma:ConditionalExpectedArrival}}\label{proof:lemma_taskArrival}
\vspace{-5pt}
Noting the expect task arrival of stable CSs is proportional to the average number of connected mobiles. Therefore, we have $\E[\Lambda|\Lambda<\mu_{\max}]= \beta^{*} \cdot  \frac{\sum_{n=1}^{    R } n \cdot \frac{\bar{N}^{n} e^{-\bar{N}}}{n!}}{1- \rho} =\bar{N}\beta^* \l(1-\frac{\Pr(N = \l\lfloor R\r\rfloor)}{1 - \rho}\r)$, where $\Pr(N = \l\lfloor R\r\rfloor) = \frac{\bar{N}^{\l\lfloor R\r\rfloor} e^{-\bar{N}}}{\l\lfloor R\r\rfloor!}$, ending the proof.

\vspace{-20pt}
\subsection{Proof of Theorem \ref{ub_Tc_heavy_VMs}}\label{proof:lemma_Tc_heavy_VMs}
\vspace{-5pt}
Substituting $\tau = 1$ into \eqref{Tcompa} and then applying $\l(\frac{\Lambda}{\mu^-(m_{\max})}\r)^{m_{\max}} \leq \frac{\Lambda}{\mu^-(m_{\max})}$ as well as  $\l(1-\frac{\Lambda}{m_{\max}\mu^-(m_{\max})} \r)^2\leq\l(1-\frac{1}{m_{\max}}\r)^2$ give an upper bound of $\mathsf{T}_{\textrm{comp}}^+(\Lambda)$ as
\begin{align}
\mathsf{T}_{\textrm{comp}}^+(\Lambda) \leq \frac{m_{\max}}{\mu^-(m_{\max})} +  \frac{\frac{\Lambda}{\mu^-(m_{\max})}}{m_{\max}!\mu^-(m_{\max})\l( 1-\frac{1}{m_{\max}} \r)^2}. \label{UBTcompa}
\end{align}
The spatial average on \eqref{UBTcompa} based on Lemma~\ref{Lemma:ConditionalExpectedArrival} gives the final result in Theorem \ref{ub_Tc_heavy_VMs}.
\newpage
\section*{Appendix II: Notation Table}
\vspace{5mm}

\begin{table}[h]
\begin{center}
\vspace{-5mm}
\begin{tabular}{|c!{\vrule width 1.5pt}l|}
\hline
Notation&Meaning\rule{0pt}{3mm}\\
\hhline{|=|=|}
$\Omega$, $\lambda_b$, $\Phi$, $\lambda_m$ & PPP of APs, density of $\Omega$, PPP of mobiles, density of $\Phi$\\
\hline
$r_0$, $\mathcal{O}(Y, r_0)$, & Offloading range, MEC service zone\\
\hline
$L$, $p$, $p_L$ & Number of slots per frame, task-generating rate at mobile, task-offloading probability\\
\hline
$\eta$, $g$, $\alpha$, $\theta$, $G$ & Transmit power of AP, fading factor, path-loss exponent, SIR threshold, spreading factor\\
\hline
$t_0$, $B$, $\ell$, $d$  & Slot length (in sec), Bandwidth per channel, number of bits per task, degradation factor of I/O interference \\
\hline
$T_0$, $\mathsf{T}_{\textrm{comm}}$, $\mathsf{T}_{\textrm{comp}}$  & Expected computation time per task, average comm-latency, average comp-latency\\
\hline
$\Lambda$, $\mu$, $m_{\max}$, $\mu_{\max}$ & Task arrival rate to CS, computation rate, maximal number of VMs, maximum computation rate\\
\hline
$\mathcal{A}_t$, $\mathcal{C}_t$,  $\mathcal{Q}_t$  &  Number of task arrival at frame $t$, number of task departure during frame $t$, the remaining tasks at frame $t$\\
\hline
$\epsilon$, $\rho$, $\bar{N}$  & Network coverage parameter, network stability parameter, expected number of mobiles connected to a AP\\
\hline
$\beta^*$, $\bar{\Lambda}^*$, $\bar{A}^*$  & Expected task generation rate, expected task arrival rate, expected number of arrival tasks per frame\\
\hline
\end{tabular}
\vspace{-10pt}
\end{center}
\caption{Summary of Notation}
\label{tab:para_system}
\vspace{-10pt}
\end{table}

\vspace{-20pt}
\section*{Appendix III: Supplementary Simulations}

 

\begin{figure}[h]
\centering
\vspace{-10pt}
\subfigure[Effect of mobile density.]{\includegraphics[width=8cm]{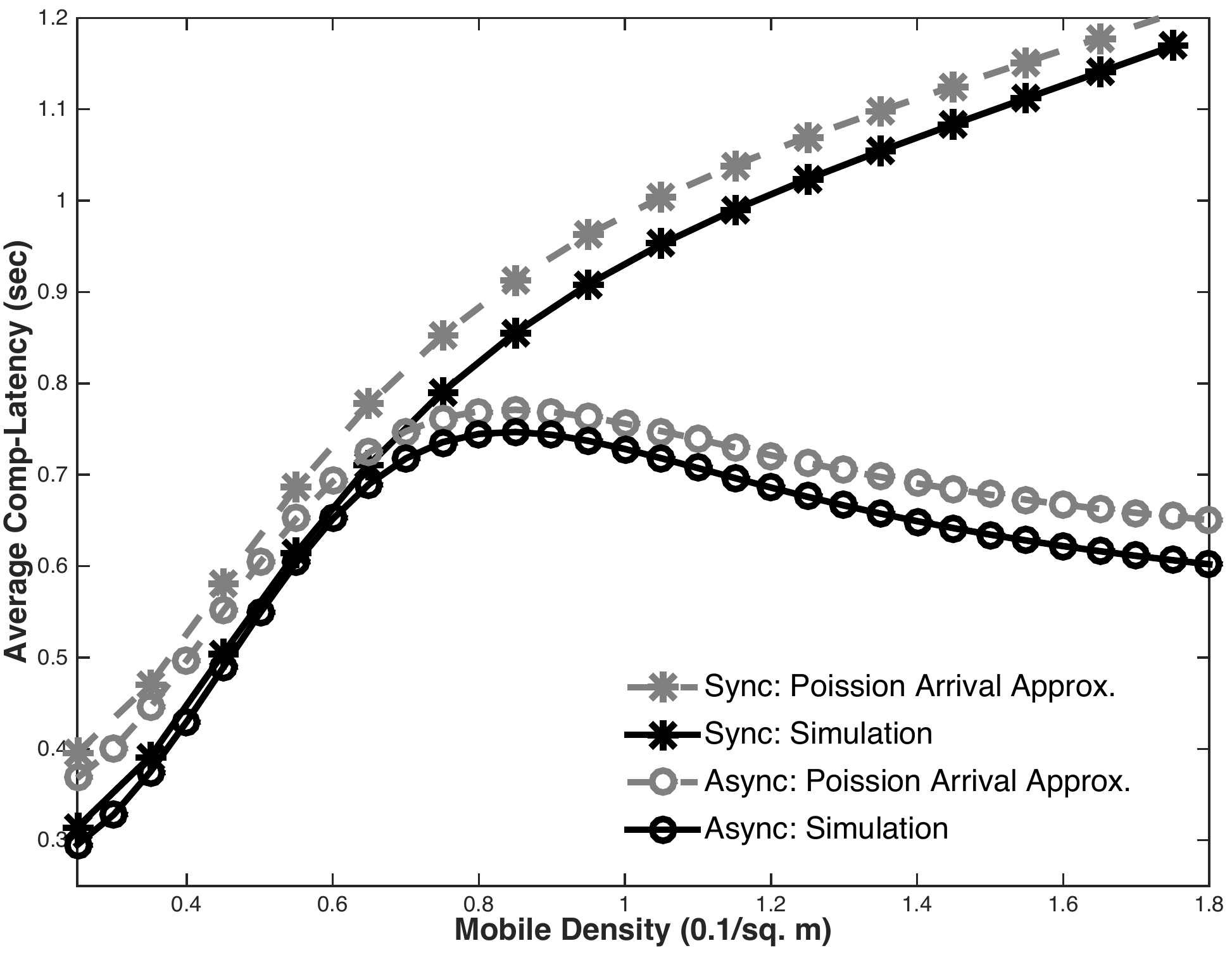}}
\subfigure[Effect of task generating rate.]{\includegraphics[width=8cm]{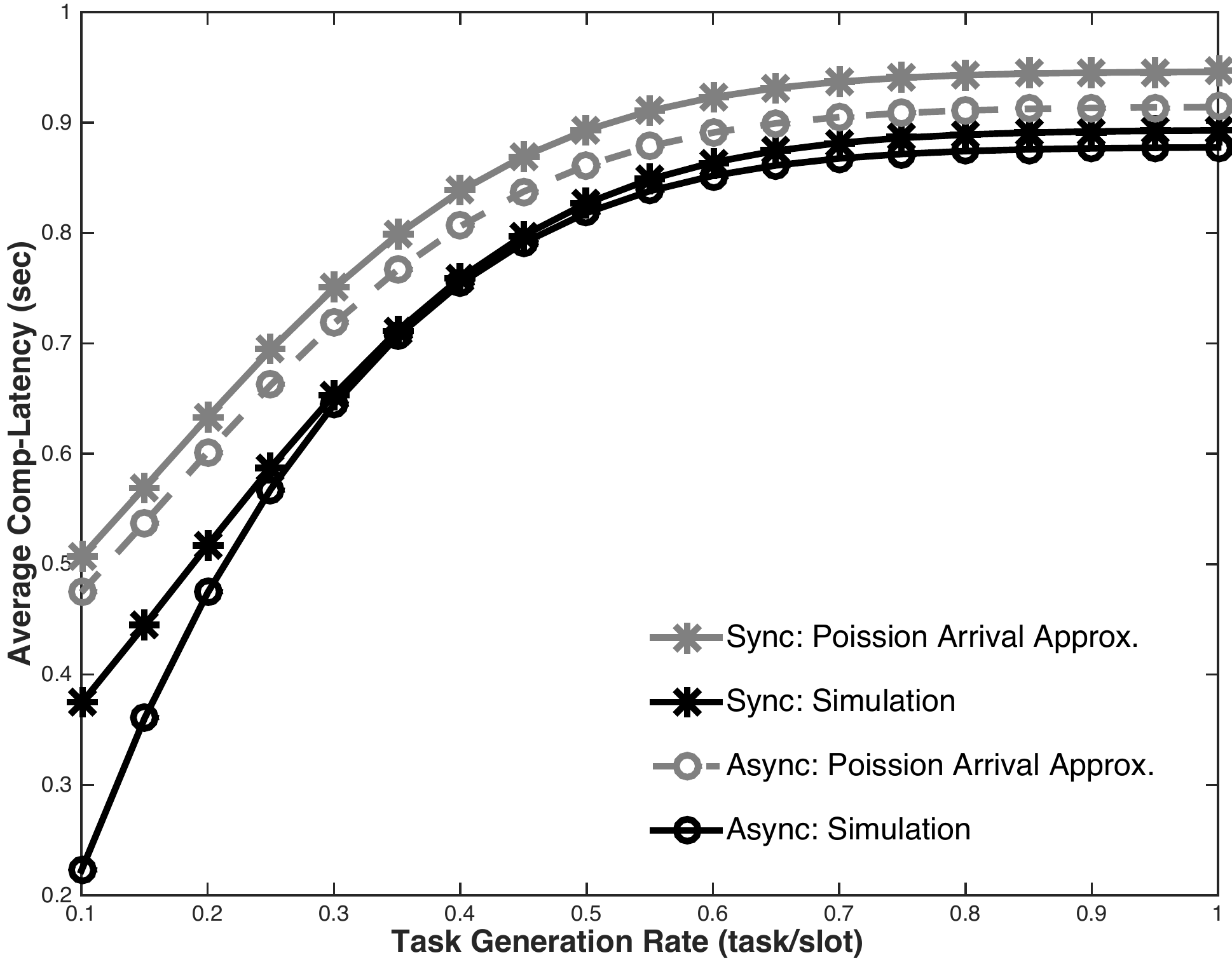}}
\vspace{-10pt}
\caption{Comparison to the Poisson Approximation in Assumption \ref{Assumption:PoissonArrival}.}
\vspace{-20pt}
\label{fig_asyLaten_arxiv}
\end{figure}

\begin{figure}[h]
\centering
\subfigure[Asynchronous offloading.]{\includegraphics[width=8cm]{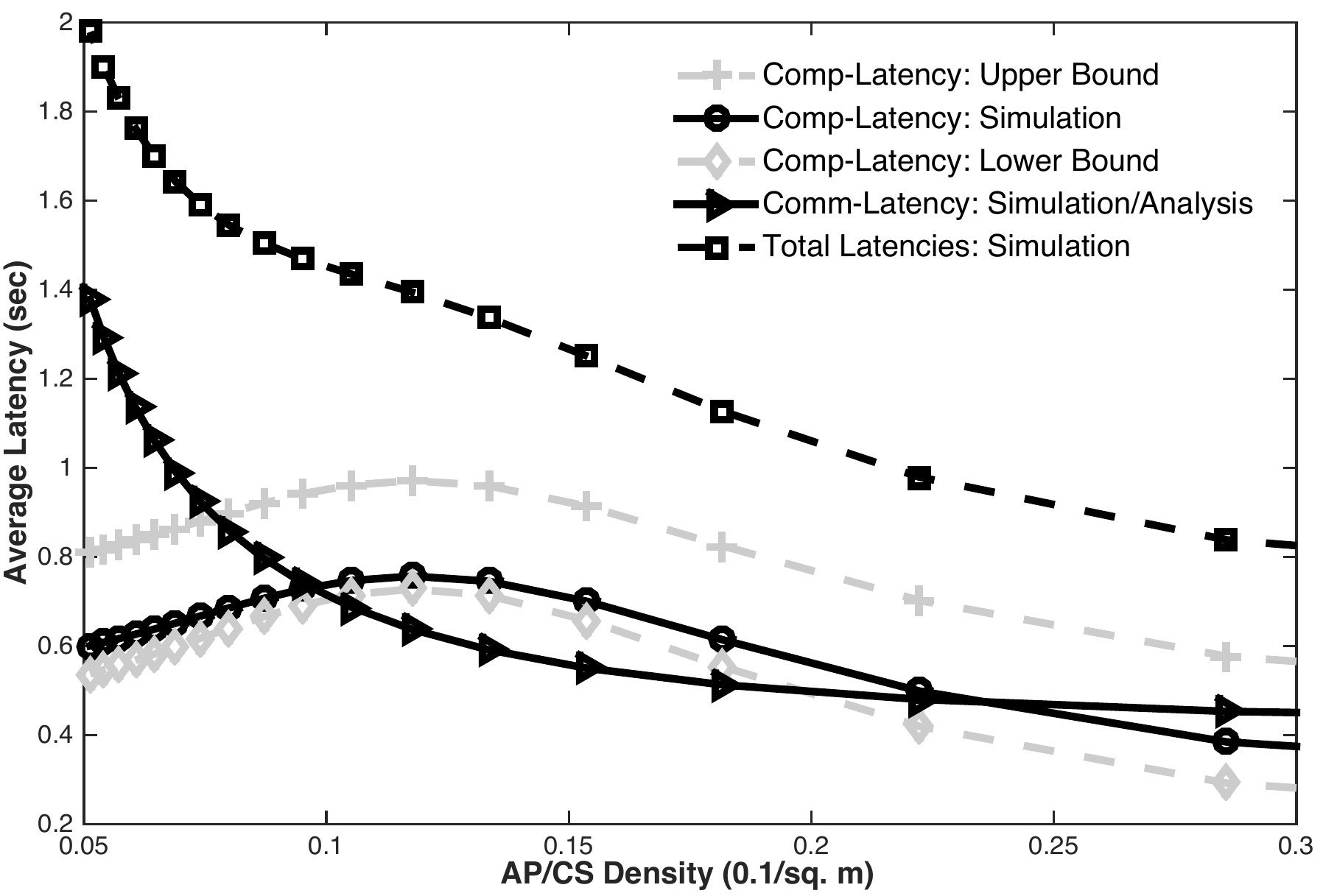}}
\subfigure[Synchronous offloading.]{\includegraphics[width=8cm]{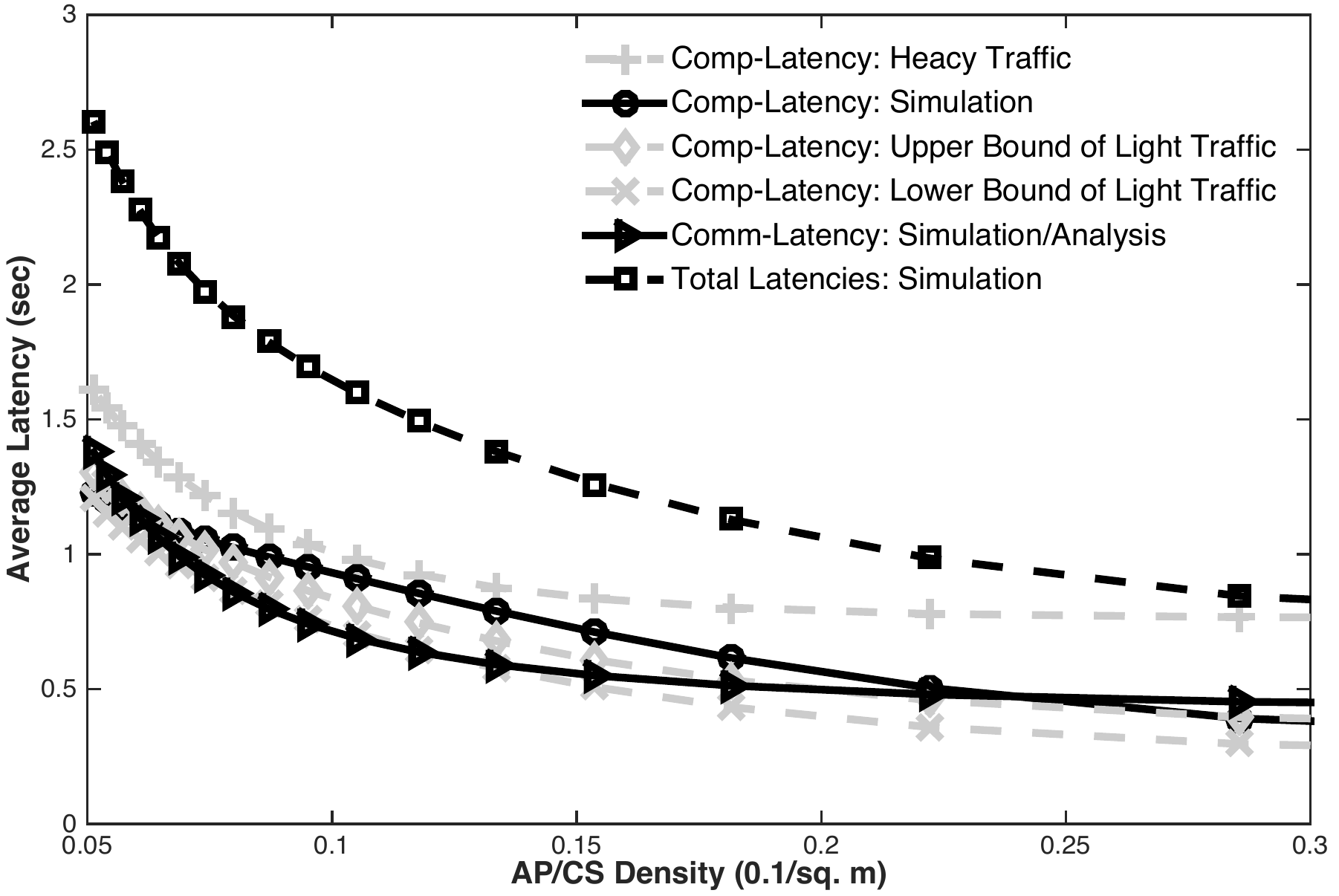}}
\caption{Effects of AP (CS) density on comm-latency and comp-latency }
\label{fig_Laten_AP}
\end{figure}

\begin{figure}[h]
\centering
\subfigure[Effect of edge-computing time per task.]{\includegraphics[width=8cm]{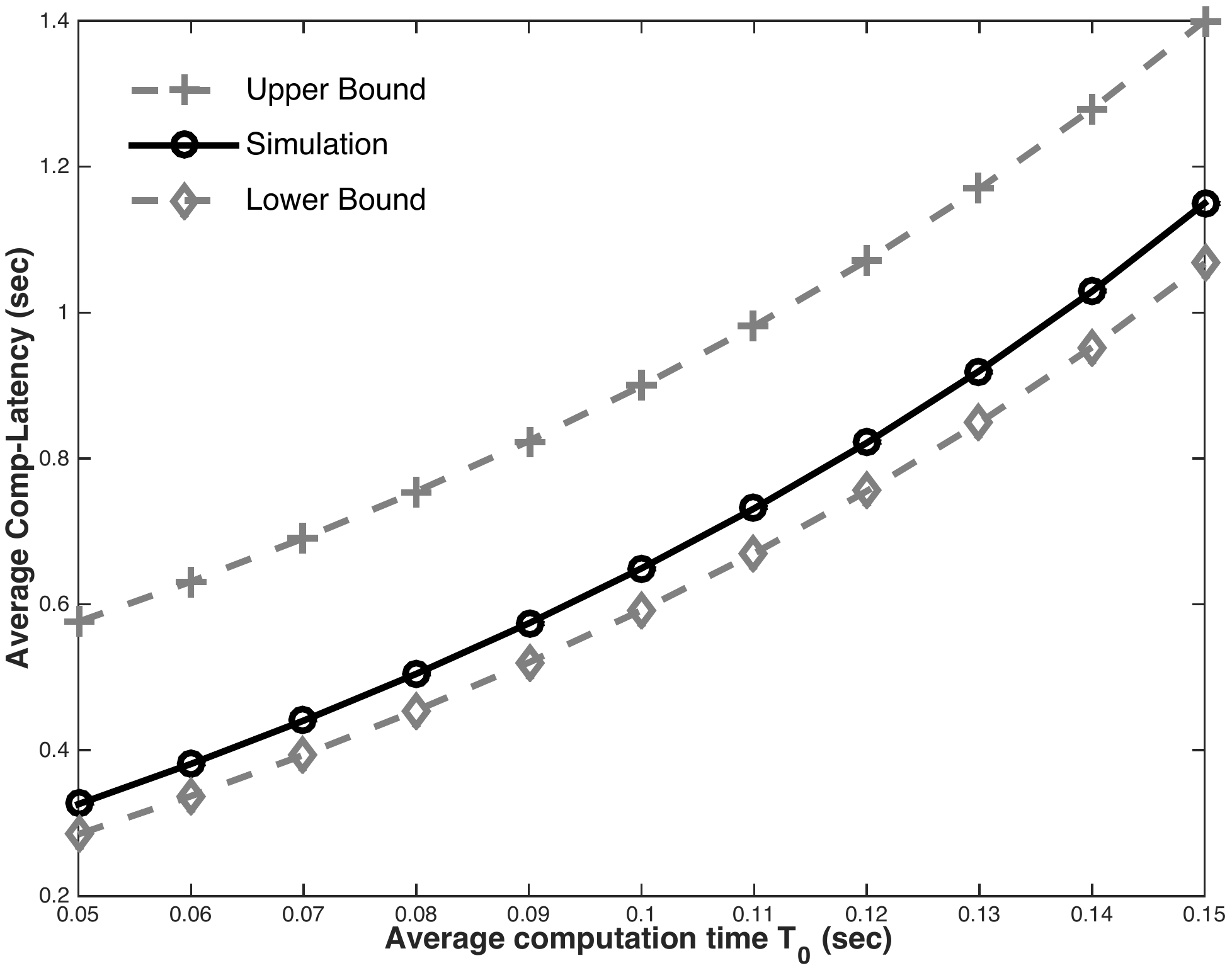}}
\subfigure[Effect of degradation factor due to I/O interference.]{\includegraphics[width=8cm]{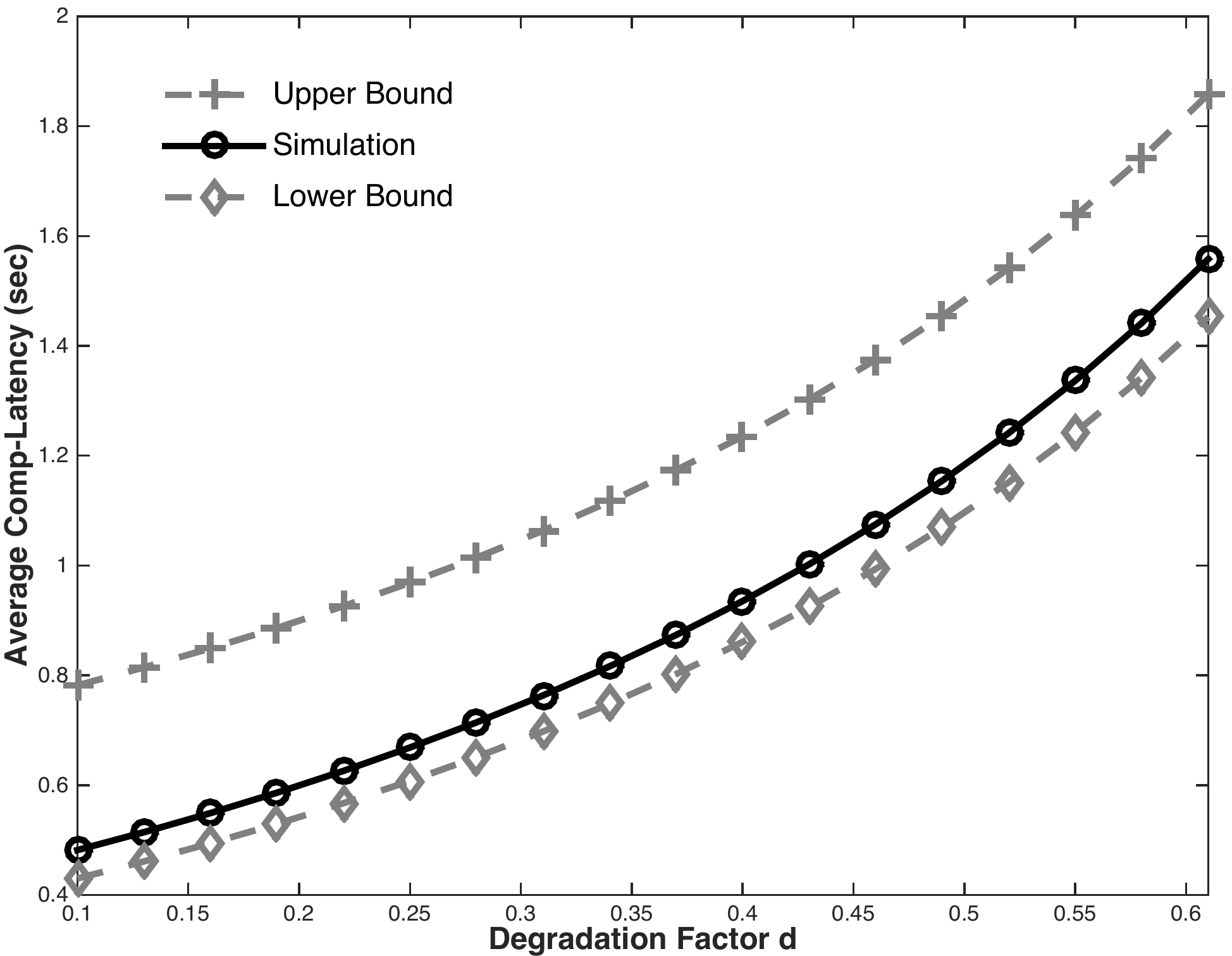}}
\vspace{-10pt}
\caption{The performance of average latencies for the asynchronous offloading case. }
\label{fig_asyLaten_d_arxiv}
\vspace{-20pt}
\end{figure}

\begin{figure}[h]
\centering
\subfigure[Effect of edge-computing time per task.]{\includegraphics[width=8cm]{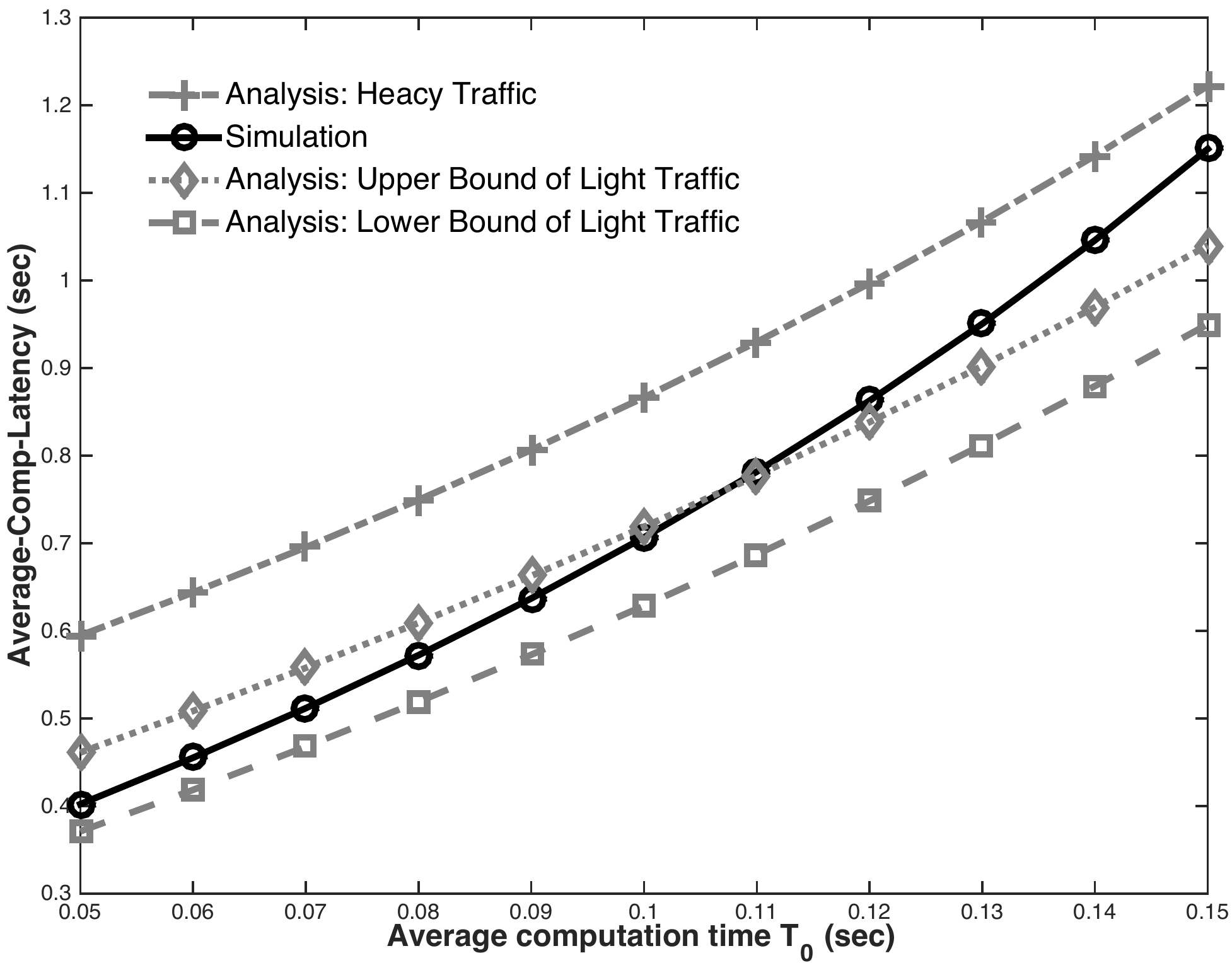}}
\subfigure[Effect of degradation factor due to I/O interference.]{\includegraphics[width=8cm]{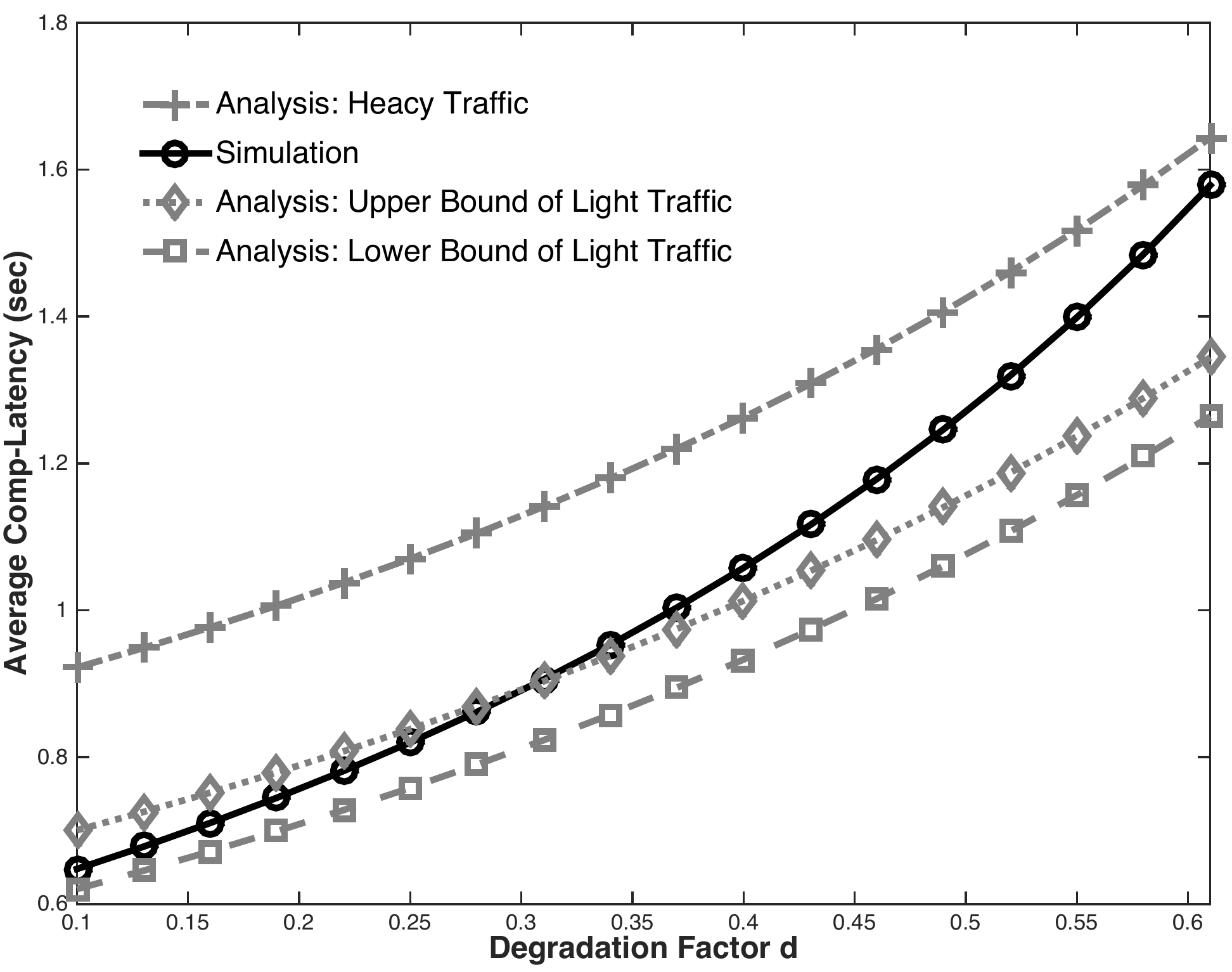}}
\vspace{-10pt}
\caption{The performance of average latencies for the synchronous offloading case. Both the light- and heavy-traffic cases are plotted.}
\vspace{-20pt}
\label{fig_synLaten_d_arxiv}
\end{figure}

\begin{figure}[h]
\centering
\subfigure[Effect of mobile density.]{\includegraphics[width=8cm]{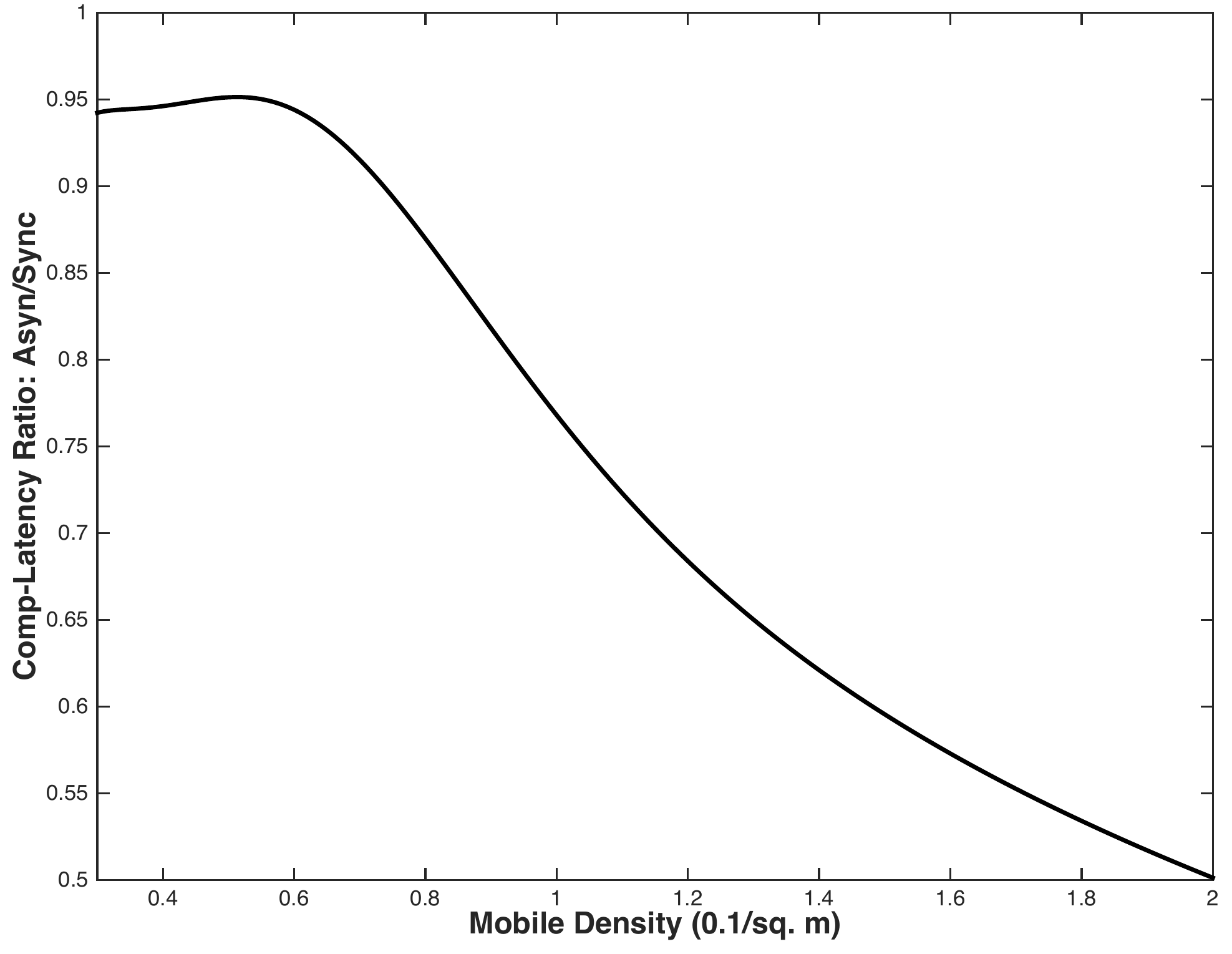}}
\subfigure[Effect of task generation rate]{\includegraphics[width=8cm]{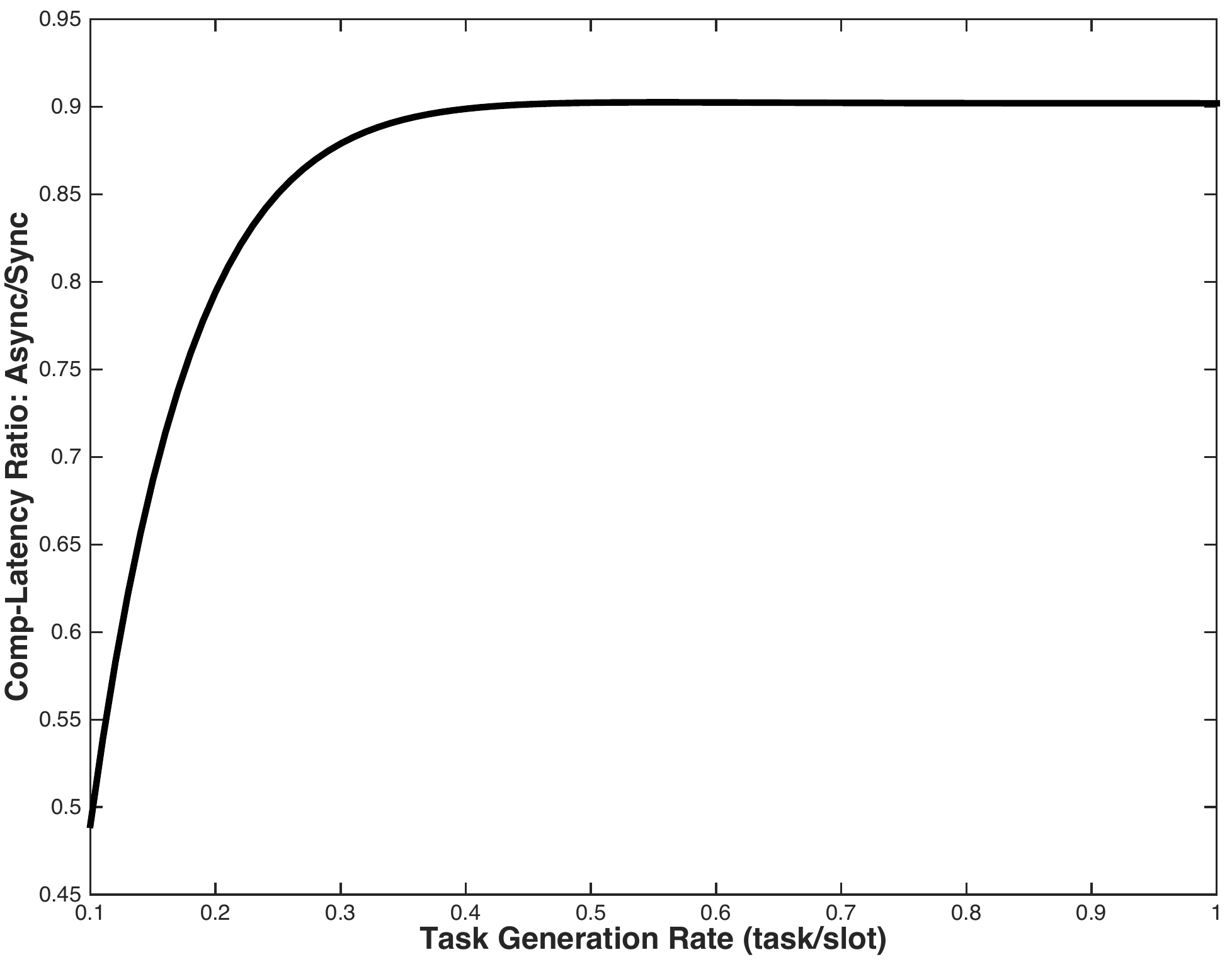}}
\vspace{-10pt}
\caption{The average comp-latency ratio: asynchronous offloading over synchronous offloading.}
\label{fig_ratio}
\vspace{-30pt}
\end{figure}

In Fig.~\ref{fig_asyLaten_arxiv}, the the comp-latency of the Poisson arrival assumption of task-arrival process is compared with that under the Poisson arrival assumption of Assumption \ref{Assumption:PoissonArrival},
showing that the tasks arrival process at AP can be well approximated to the Poisson task arrival assumption. 

Fig.~\ref{fig_Laten_AP} represents the effects of AP density, which is shown that the same observations are made in the case of decreasing mobile density. 


Figs.~\ref{fig_asyLaten_d_arxiv} and \ref{fig_synLaten_d_arxiv}  show the effects of the computation capability of CS,  the computation time $T_0$ and the degradation factor $d$, on the comp-latency for asynchronous and synchronous offloading cases, respectively. Both of the comp-latencies increase exponentially when the edge-computing capability worse, i.e., $T_0$ or $d$ becomes larger, which agrees with the intuition.


Finally, we plot the ratio between the comp-latency of asynchronous and synchronous offloading cases in Fig.~\ref{fig_ratio}, which are always less than one because the comp-latency of the asynchronous offloading is always smaller than the synchronous counterpart. Specifically, in Fig.~\ref{fig_ratio}(a), the ratio first stays constant when mobiles is sparse and then decreases with mobile density grows. As mobiles becomes denser, 
the number of offloading tasks at the beginning of each frame increases, resulting in severe I/O interference than the asynchronous counterpart of which the task arrivals are distributed over frame. 
 On the other hand, in Fig.~\ref{fig_ratio}(b), it is observed that the ratio grows and then converges to a constant when $p$ increases. In other words, the gap of comp-latency between two offloading cases is becoming smaller when the task arrival becomes heaver, aligning with the discussion in Remark~\ref{remark:ComparingAsyncHeavy}.

\vspace{-18pt}

\bibliographystyle{ieeetr}
\bibliography{BibDesk_File}

\end{document}